\documentclass{aastex63}

\usepackage{url}

\received{November. 21, 2020}
\revised{January. 28, 2021}
\accepted{\today}
\submitjournal{ApJS}

\shorttitle{O-type stars found with LAMOST}
\shortauthors{G.-W. Li}
\begin{document}

\title{Galactic O-type stars in LAMOST data}

\correspondingauthor{Guang-Wei Li}
\email{lgw@bao.ac.cn}

\author[0000-0001-7515-6307]{Guang-Wei Li}
\affiliation{Key laboratory of Space Astronomy and Technology, National Astronomical Observatories, Chinese Academy of Sciences, 
Beijing 100101, China}



\begin{abstract}
This paper reports 209 O-type stars found with LAMOST. All 135 new O-type stars discovered so far with LAMOST so far are given. Among them, 94 stars are firstly presented in this sample. There are 1 Iafpe star, 5 Onfp stars, 12 Oe stars, 1 Ofc stars, 3 ON stars, 16 double-lined spectroscopic binaries, and 33 single-lined spectroscopic binaries. All O-type stars are determined based on LAMOST low-resolution spectra ($R \sim 1,800$), with their LAMOST median-resolution spectra ($R\sim 7,500$) as supplements. 
\end{abstract}


\section{Introduction} \label{sec:intro}

O-type stars are very rare and short-lived, but they play important roles in many astronomical fields. They illuminate the galaxy where they live. They produced tremendous amounts of ionizing photons that punctured the neutral gas and made the universe transparent in the Epoch of Reionization\citep{hai97,loeb01}. Their stellar winds and powerful explosions when they die are drivers of gas in galaxies \citep{rog13}, while their ejecta can enrich surrounding gas where new stars are born. They are thought to be the progenitors of many core-collapse supernovae (CCSNe), and long gamma-ray bursts (LGRBs) \citep{woo06,langer12}, and also responsible for some gravitational-wave events \citep{de16}. Besides, massive stars are good tools to study the structure of the Galactic plane \citep{xu18,chen19,cheng19}. 

\citet{sana12} presented that most O-type are born in binary systems, and about three-quarters of O-type stars would undergo interactions. Thus, multiple observations for these O-type stars from different epochs are necessary to understand their interaction processes and nature. As a result, the compilation of Galactic massive stars is a fundamental  work. \citet{reed03}  gave a widely used catalog of OB stars based on optical photometry. Later, \citet{mai04} gave a catalog of 378 O-type stars based on optical spectroscopy. Then, the Galactic O-Star Spectroscopic Survey (GOSSS)  \citep{mai11} presented 594 O-type stars by a uniform spectral classification criteria \citep{sota11, sota14, mai16}. Recently, \cite{liu19} presented the largest OB catalog from Guoshoujing Telescope (the Large Sky Area Multi-Object Fiber Spectroscopic Telescope, LAMOST) database, and \cite{roman19} also gave a catalog of 27 new O-type stars based on the LAMOST database. Beyond the Galaxy, 213 O-type stars in 30 Doradus were given by \citet{wal14}. 
 
In this paper, I compile 209 O-type stars found in the LAMOST database until the end of 2018. 

\section{Data} \label{sec:Data}
I used the method given in \citet{li20a} to select O-type stars in the LAMOST low-resolution spectral data ($R\sim 1,800$) between 2011 and 2018, and LAMOST medium-resolution spectra ($R \sim 7,500$) are also used, if any. Information about LAMOST low- and medium-resolution spectra can be referred to \citet{wang96}, \citet{su04}, \citet{cui12}, \citet{luo12}, \citet{zhao12} and \citet{liu20}. Each star is assigned with the name of "LAMOST-O xxx", where "xxx" is a number ascending with the star's galactic longitude. All O-type stars found in LAMOST databases, with their names recommended by Simbad, sky positions, and spectral classifications, are tabulated in Table \ref{tab:lamosto}.  The LAMOST low- and medium-resolution spectra shown in this paper are provided in the China-VO PaperData repository: \href{https://nadc.china-vo.org/registry/paperdata/101049}{doi:10.12149/101049}

\section{Spectral classification method} \label{sec:speclass}
In this section, a brief introduction to the spectral classification scheme is given. The detailed spectral classification criteria for O-type stars are from \citet{sota11,sota14} and \citet{mai16}, while spectral type and luminosity sequences are shown in Figures 3-11 in \citet{sota11}.

\begin{figure*}
\center
%\begin{center}
\includegraphics[angle=0, scale=0.8]{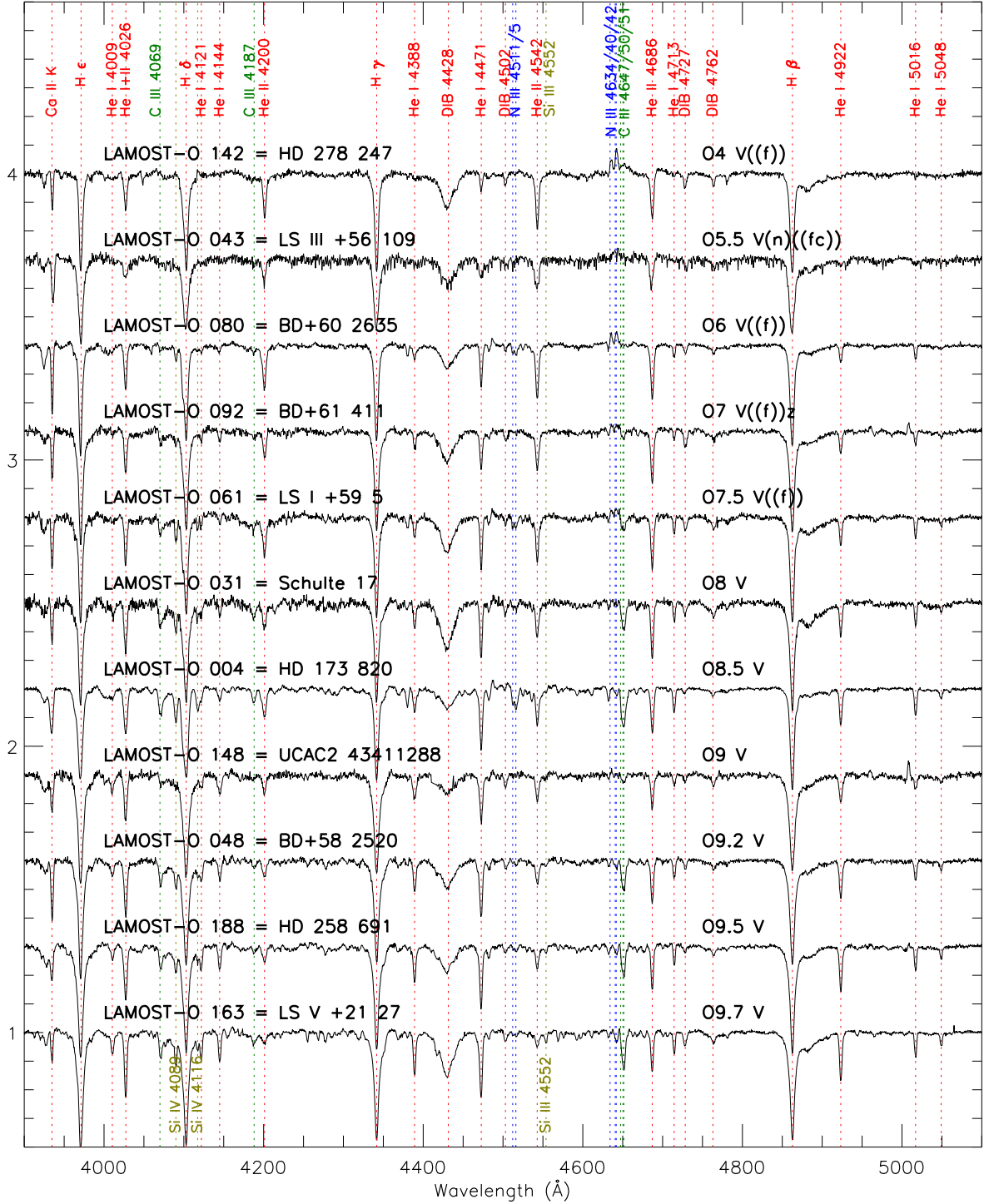}
%\plotone{iafpe.eps}
\caption{A spectral type sequence of LAMOST O dwarfs.}
\label{fig:spt}
%\end{center}
\end{figure*}

\begin{figure*}
\center
%\begin{center}
\includegraphics[angle=0, scale=0.8]{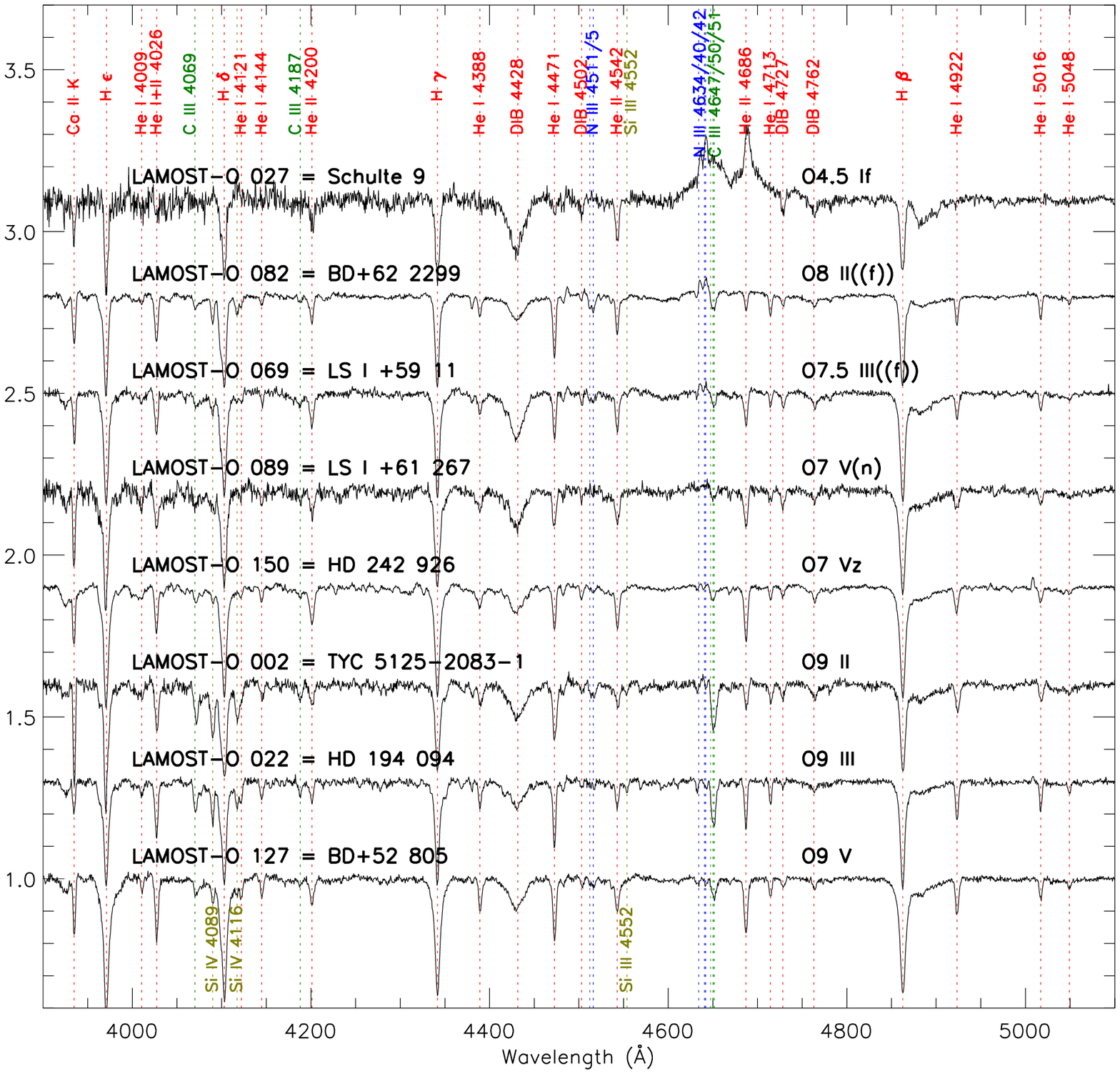}
%\plotone{iafpe.eps}
\caption{Luminosity sequences of LAMOST O-type stars. The top five spectra are for earlier stars, while the bottom three spectra for later stars.}
\label{fig:lum}
%\end{center}
\end{figure*}

\begin{enumerate}
\item The spectral type is primarily based on the ratio of He II $\lambda$4542/He I $\lambda$4471. The later the spectral type, the smaller the ratio, and is $\sim$ 1 at O7. For types equal to or later than O8, He II $\lambda$4542/He I $\lambda$4388 and He I $\lambda$4388/He I $\lambda$4144 become very useful to determine spectral type. At O9, both are $\sim$ 1. Moreover, the ratio of Si\,III $\lambda$4552/He II $\lambda$4542 becomes useful for determining spectral types later than O9, and $\sim$1, at O9.7 (see Table 3 in \citet{sota14}). A spectral type sequence of LAMOST O dwarfs are shown in Figure \ref{fig:spt}.
\item For stars earlier than O9, the luminosity classification is based on the features of He II $\lambda$4686 and  N III $\lambda$4634-4640-4642, which define the 'Of' phenomenon: ((f)) means weak emission of the N III line, while there is strong absorption of the He II line; (f) means medium N III emission, while the He II absorption is weak; f means both lines are in strong emission. The stronger the Of effect, the higher the luminosity (see Table 2 in \citet{sota14}). 
A luminosity sequence for O-type stars earlier than O9 is shown by the top five spectra in Figure. \ref{fig:lum}.
\item For stars equal to or less than O9, the luminosity classification is based on the ratios of He II $\lambda$4686/He I $\lambda$4713 and He I $\lambda$4026/Si IV $\lambda$4089. The the lower the ratios, the higher the luminosity (see Table 6 in \citet{sota11}). A luminosity sequence for O9 is shown by the bottom three spectra in Figure. \ref{fig:lum}.
\item The suffix $z$ is assigned to stars with the ratio of  $\mathbf{EW}$(He II $\lambda$4686) / Max[$\mathbf{EW}$(He II $\lambda$4686), $\mathbf{EW}$(He II $\lambda$4686)] $>$ 1.1, which are supposed to be very near the zero-age main sequence \citep{sab14,arias16}.
\item As in \citet{sota11}, (n), n and nn are used to indicate the rotational broadening, with roughly projected rotational velocities of 200 km\,s$^{-1}$, 300 km\,s$^{-1}$, and 400 km\,s$^{-1}$, respectively.
\end{enumerate}

The MK spectral classification is to compare the spectrum under study to standard spectra. It is impossible for these standard spectra to cover all kinds of spectra. Thus, compromises are needed. Moreover, the signal-to-noise ratio, rotational effect, stellar wind, metallicity, and even spectral resolution influence the classification. Thus, the spectral classification is something of an art.
\par

So far, there are still no widely accepted quantitative spectral classification criteria. \citet{mar18} gave a quantitative spectral classification based on the equivalent widths (EWs) of some lines, but cannot assign an exact spectral type to a star with only these quantities. Moreover, both the quality and resolution of LAMOST spectra are low. As a result, the classic MK spectral classification method is used. Specifically, all O-type stars in this paper are assigned spectral types by comparing with standard O-type stars in Table 2 in \citet{mai16}.

\section{Result} \label{sec:result}
\subsection{Iafpe} \label{subsec:iafpe}

O Iafpe stars are Of stars, with He\,I $\lambda$4471 showing P Cygni profiles \citep{sota14}. 

\begin{figure*}
\center
%\begin{center}
\includegraphics[angle=0, scale=0.8]{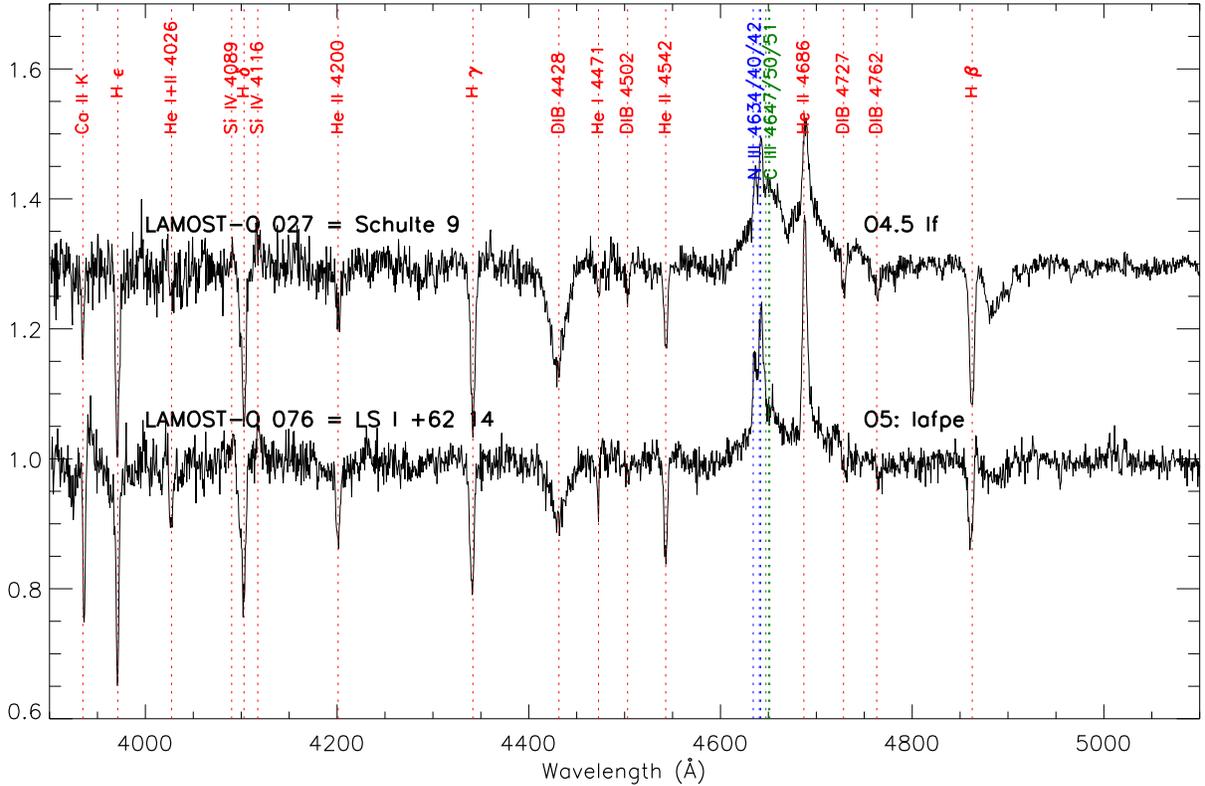}
%\plotone{iafpe.eps}
\caption{Spectra of LAMOST-O 027 and 076.}
\label{fig:iafpe}
%\end{center}
\end{figure*}

\emph{LAMOST-O 027 = Schulte 9}. This star is also in GOSSS \citep{sota14}, with a spectral type of O4.5 If. By comparing two spectra given by LAMOST (see Figure \ref{fig:iafpe}) and GOSSS, respectively, this star is an SB1 ($\Delta V \sim 44$ km\,s$^{-1}$) and also shows variations of the emission lines, but its spectral type is unchanged. In high-quality spectra, it is an SB2 \citep{naz12}.

\emph{LAMOST-O 076 = LS I +62\,14}. In \citet{skiff14}, its spectral type is B0 or OBe. Thus, it may be a new O-type star. Its spectrum is very similar to that of LAMOST-O 027 in Figure \ref{fig:iafpe}: He II $\lambda$4686 and N III $\lambda$4634-40-42 are both in strong emissions. However, their He I $\lambda$4471 and H $\beta$ are different.

In the spectrum of LAMOST-O 076, the He I $\lambda$4471 absorption is very sharp, which may be infilled by emission produced by the material in the stellar wind. Moreover, there may be  emission in the blue wings of  He I $\lambda$4471 and H $\beta$, though weak, so  He I $\lambda$4471 and H $\beta$ may be P Cygni profiles. As a result, LAMOST-O 076 is an O5: Iafpe candidate.

\subsection{Onfp} \label{subsec:onfp}
\begin{figure*}
\center
%\begin{center}
\includegraphics[angle=0, scale=0.8]{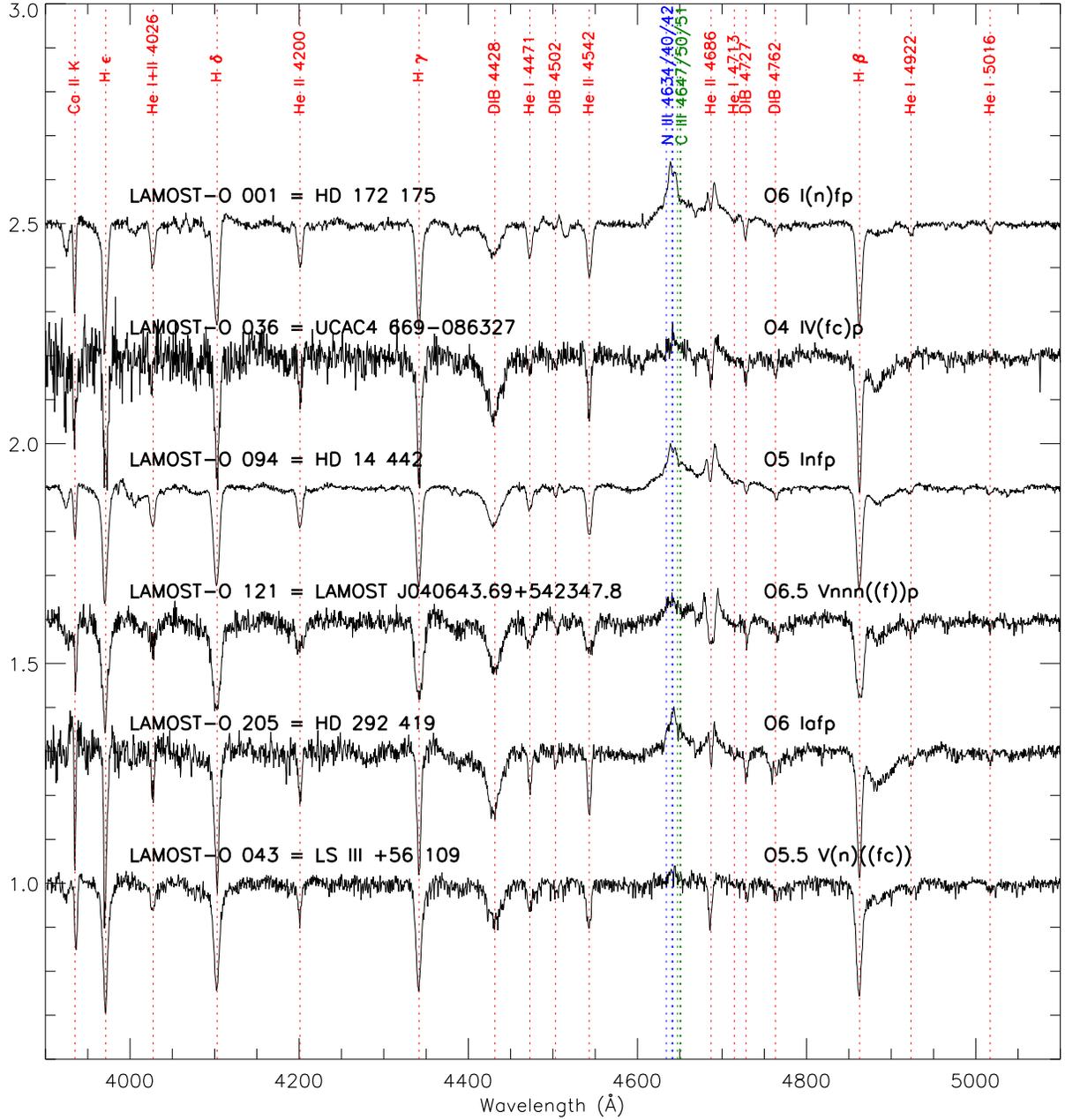}
%\plotone{iafpe.eps}
\caption{Spectra of Onfp stars (LAMOST-O 001, 036, 094, 121, and 205) and the Ofc star (LAMOST-O 043).}
\label{fig:onfp}
%\end{center}
\end{figure*}

The Onfp category consists of fast-rotating Of stars that have absorption reversals in its He\,II $\lambda$4686 emissions \citep{sota14}. All LAMOST spectra of Onfp stars are shown in Figure \ref{fig:onfp}.

\emph{LAMOST-O 001 = HD 172\,175}. This star is also in GOSSS. By comparing two spectra given by LAMOST and GOSSS \citep{sota11}, respectively, the ratio of He II $\lambda$4200/He I $+$ II $\lambda$4026 is $\sim$ 1 in the LAMOST spectrum, a little higher than that in the GOSSS spectrum, which results in a little earlier spectral type in the LAMOST spectrum, O6 I(n)fp.

\emph{LAMOST-O 036 = UCAC4 669-086327}. In \citet{roman19}, this star is O4 IV(f). In the LAMOST spectrum, its C\,III $\lambda$4647-50-51 emission is also strong enough to be assigned to a suffix of (c). Moreover, its He II $\lambda$4686 has a weak emission wing. As a result, it is an O4 IV(fc)p.

\emph{LAMOST-O 094 = HD 14\,442}.  Comparing two spectra given by LAMOST and GOSSS \citep{sota11}, it is an SB1 star ($\Delta V \sim 72$ km\,s$^{-1}$), with a variable He\,II $\lambda$4686 profile.

\emph{LAMOST-O 121 = LAMOST J040643.69+542347.8}. This star is the fastest rotator in the Galaxy with a $v_e\sin i \sim 540$ km\,s$^{-1}$. It is also a runaway, which implies binary origin. For detailed information, refer to \citet{li20b}.

\emph{LAMOST-O 205 = HD 292\,419}. It is a B-type star in \citet{reed03}, but its LAMOST spectrum is O6 Iafp.

\subsection{Ofc} \label{subsec:ofc}

The Ofc category consists of normal spectra with the emission intensity of C III $\lambda$4650 similar to that of N III $\lambda$4634 \citep{sota14}, and  peaks at O5 for all luminosity classes. There is only one Ofc star in LAMOST data as shown in Figure \ref{fig:onfp}. 

\emph{LAMOST-O 043 = LS III +56 109}. This star is an SB2 in \citet{mai16}. Comparing its spectra from LAMOST and GOSSS, there is a $\Delta V$ of $\sim 155$ km\,s$^{-1}$.

\subsection{Oe} \label{subsec:oe}
\begin{figure*}
\center
%\begin{center}
\includegraphics[angle=0, scale=0.8]{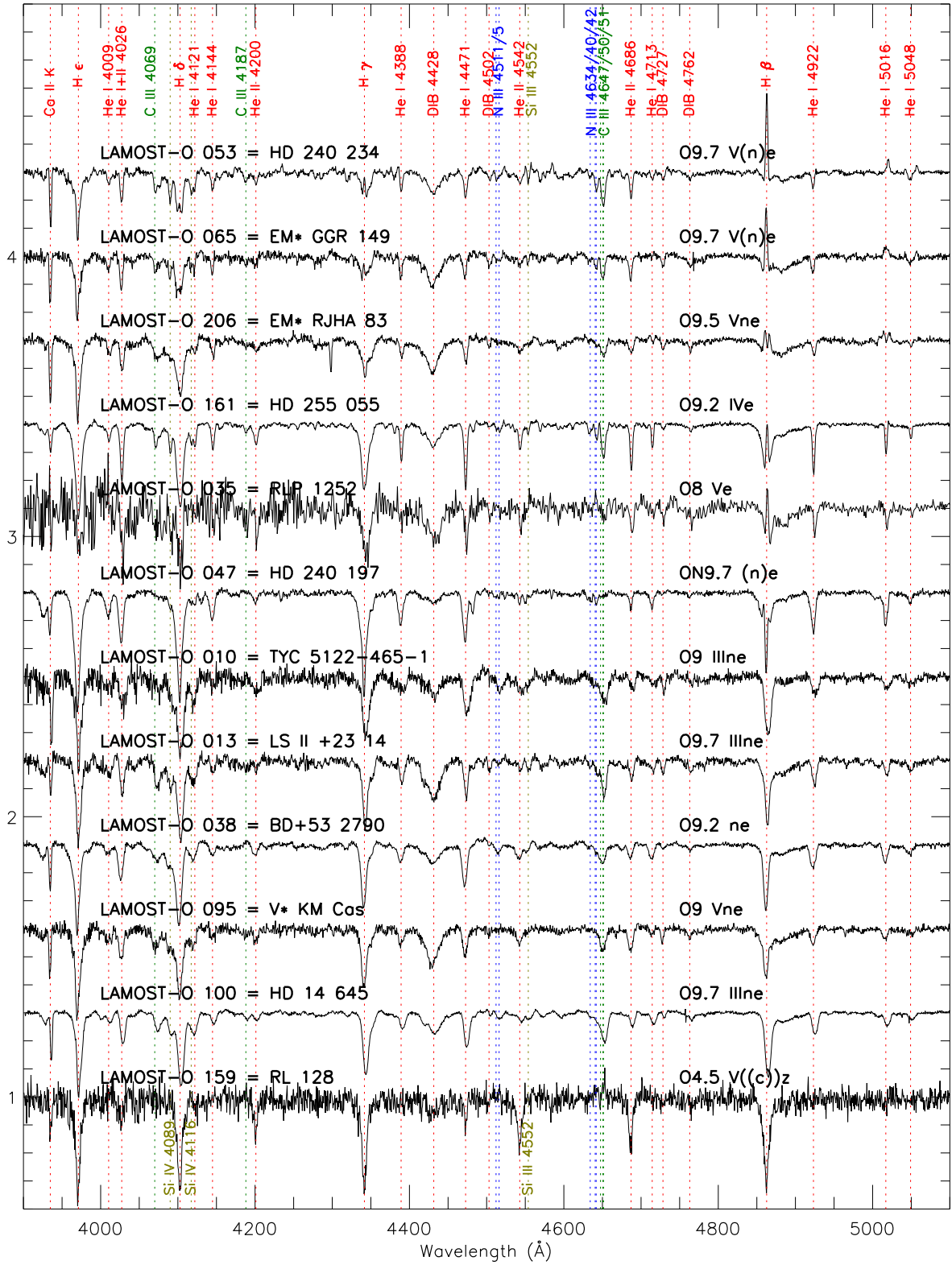}
%\plotone{iafpe.eps}
\caption{Spectra of Oe stars (LAMOST-O 010, 013, 035, 038, 047, 053, 065, 095, 100, 159, 161, and 206).}
\label{fig:oe}
%\end{center}
\end{figure*}

The Oe category was defined by \citet{conti74} and is thought to be the hotter counterpart of the classic Be category \citep{riv13}. Until 2018, there were only 13 Oe stars in the Galaxy \citep{li18}. Here, I present another 12 Oe-type stars found with LAMOST (see Figure \ref{fig:oe}). Among them, LAMOST-O 013 = LS II +23 14,  LAMOST-O 065 = EM* GGR 149, LAMOST-O 095 = V* KM Cas, LAMOST-O 159 = RL 128, LAMOST-O 161 = HD 255\,055, and LAMOST-O 206 = EM* RJHA 83 = TYC 4801-17-1 had been discussed in \citet{li18}.  

\begin{figure*}
\center
%\begin{center}
\includegraphics[angle=0, scale=0.8]{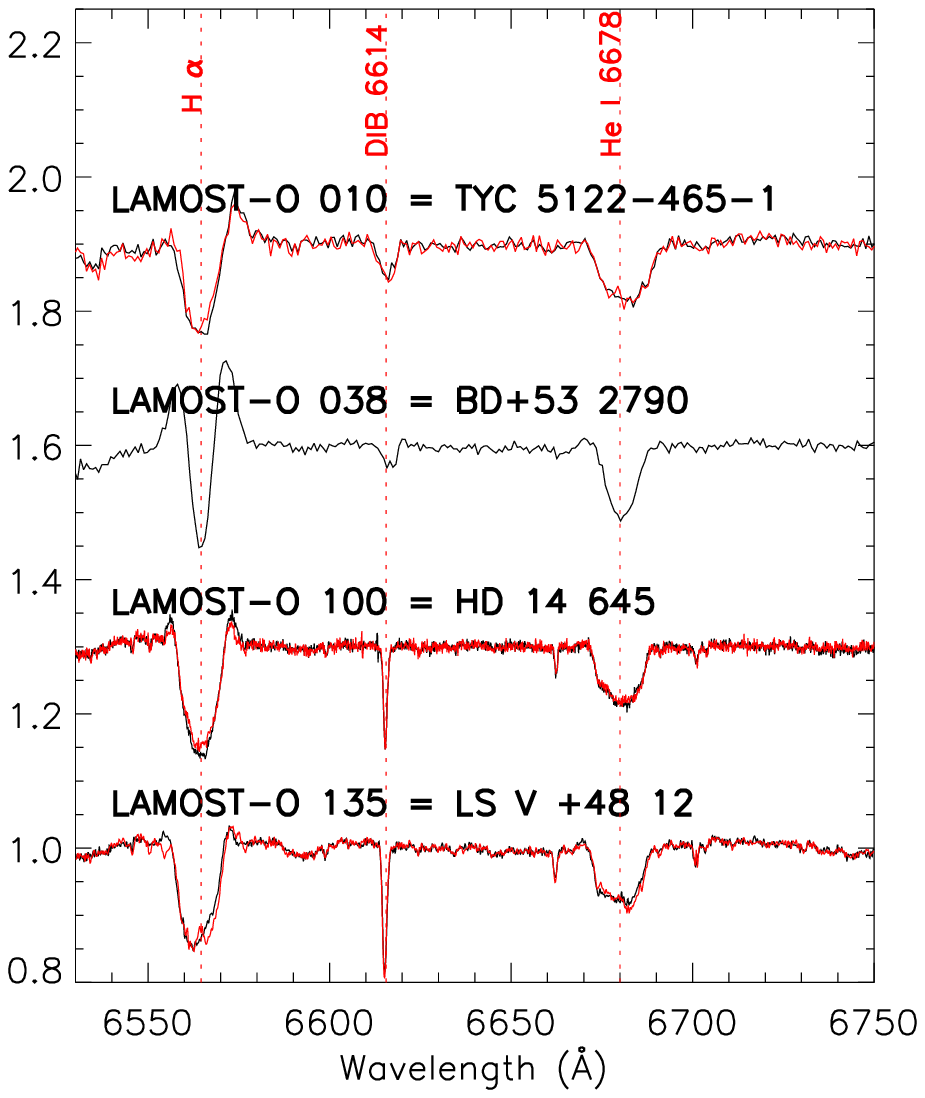}
%\plotone{iafpe.eps}
\caption{LAMOST median-resolution spectra of Oe stars (LAMOST-O 010, 038 and 100) and the SB2 LAMOST-O 135. Two spectra of a star from different epochs are shown in black and red spectra, respectively. }
\label{fig:oe_ha}
%\end{center}
\end{figure*}

\emph{LAMOST-O 010 = TYC 5122-465-1}. This star has no spectral information in Simbad, and thus may be a new O-type star. Its spectrum in Figure \ref{fig:oe} shows it is an O9 IIIn star, while emissions in the blue wings of its H$\alpha$ lines in two LAMOST low-resolution spectra (see Figure \ref{fig:oe_ha}) imply it may be an Oe star.  

\emph{LAMOST-O 035 = RLP 1252}. In \citet{roman19}, its spectral type is O8 III(f)e, but the LAMOST spectrum shows that He II $\lambda$4686 is much stronger than He I $\lambda$4713, so it is a dwarf. Moreover, its N III $\lambda$4634-40-42 is not easily recognized from noise. Thus, it is O8 Ve.

\emph{LAMOST-O 038 = BD+53 2790}. This star is an X-binary \citep{rib06}, but \citet{blay06} suggested it is a peculiar single O-type star. Its H$\alpha$ profile in Figure \ref{fig:oe_ha} indicates that it is an Oe shell-like star. However, the ratio of He II $\lambda$4686/He I $\lambda$4713 indicates that its luminosity is II, but its weak Si\,IV $\lambda$4089 and 4116 indicate a lower luminosity. Thus its luminosity is unclear. 

\emph{LAMOST-O 047 = HD 240\,197}. In Simbad, its spectral type is B5 \citep{cannon93}. Its H$\alpha$ and H$\beta$ profiles in the LAMOST spectrum suggest it is an Oe star. Moreover, its C\,III lines have almost disappeared, while the N\,III $\lambda$4634-40-42 absorption is very strong. The ratio of He II $\lambda$4686/He I $\lambda$4713 indicates that its luminosity is II, but its weak Si\,IV $\lambda$4089 and 4116 indicate it is a dwarf. Thus, its luminosity is unclear. As a result, it is assigned ON9.7 (n)e. 
\par
This star may be a pole-on, rapidly rotating star. The rapid rotation induces chemically homogeneous evolution, which can take C and H on the stellar surface into the core, while bringing the N produced by the CNO-burning cycle in the core back to the surface \citep{mae00}. The rapid rotation also lifts the material around the equator lifted to form a circumstellar disk, where emission lines are produced \citep{riv13}. The strong Mg II $\lambda$4481 also implies circumstellar material. As a result, we see an Oe star with N enrichment, but without C.

\emph{LAMOST-O 053 = HD 240\,234}. In Simbad, its spectral type is B0 \citep{bro53}, but the LAMOST spectrum clearly shows it is an O9.7 V(n)e star.

\emph{LAMOST-O 065 = EM* GGR 149}. It was classified as O9.7 IVn in \citet{li18}. But by comparing the standard stars in \citet{mai16}, it is more like a dwarf. Moreover, it should be a slower rotator compared to other rotation stars. As a result, it is assigned O9.7 V(n)e.

\emph{LAMOST-O 100 = HD 14\,645}. In Simbad, its spectral type is B0 IV:nn \citep{hilt56s}. According to the latest GOSSS criteria, it is an O9.7 star. For the same reason as LAMOST-O 038, its luminosity is unclear. The H$\alpha$ profiles in its two LAMOST medium-resolution spectra have variable emission wings (see Figure \ref{fig:oe_ha}), which suggest it is also an Oe star. As a result, it is O9.7 ne.

\emph{LAMOST-O 159 = RL 128}. This star is a Kepler K2 star, with a period of 5.03 days \citep{buy15}. Its H$\alpha$ emission is variable, and Ca II triplet emission also appears sometimes \citep{li18}. It is very dim, $V = 15.11$ mag, but near, $\varpi = 0.1752 \pm 0.0316$ mas \citep{gaia2020}, that is $\sim$ 5.7 kpc. If there is no extinction, the absolute magnitude $M_V \sim 1.33$ mag, which means that it is an sdO. In fact, it is very blue ($U - B = -0.9$ mag) \citep{mer87}, which implies extinction should be $\ll 6$ mag \citep{wal02}. As a result, this star may be a binary, and the primary is a sdO, which is accreting the material from the secondary. Because of the low qualitiy of LAMOST spectra, the GOSSS spectral type with a suffix 'e' is adopted: O4.5 V((c))ze. 

\emph{LAMOST-O 161 = HD 255\,055}. Its spectral type was O9.5 IIIe in \citet{li18}, but here it is revised to O9.2 IVe.

\subsection{ON} \label{subsec:on}
\begin{figure*}
\center
\includegraphics[angle=0, scale=0.8]{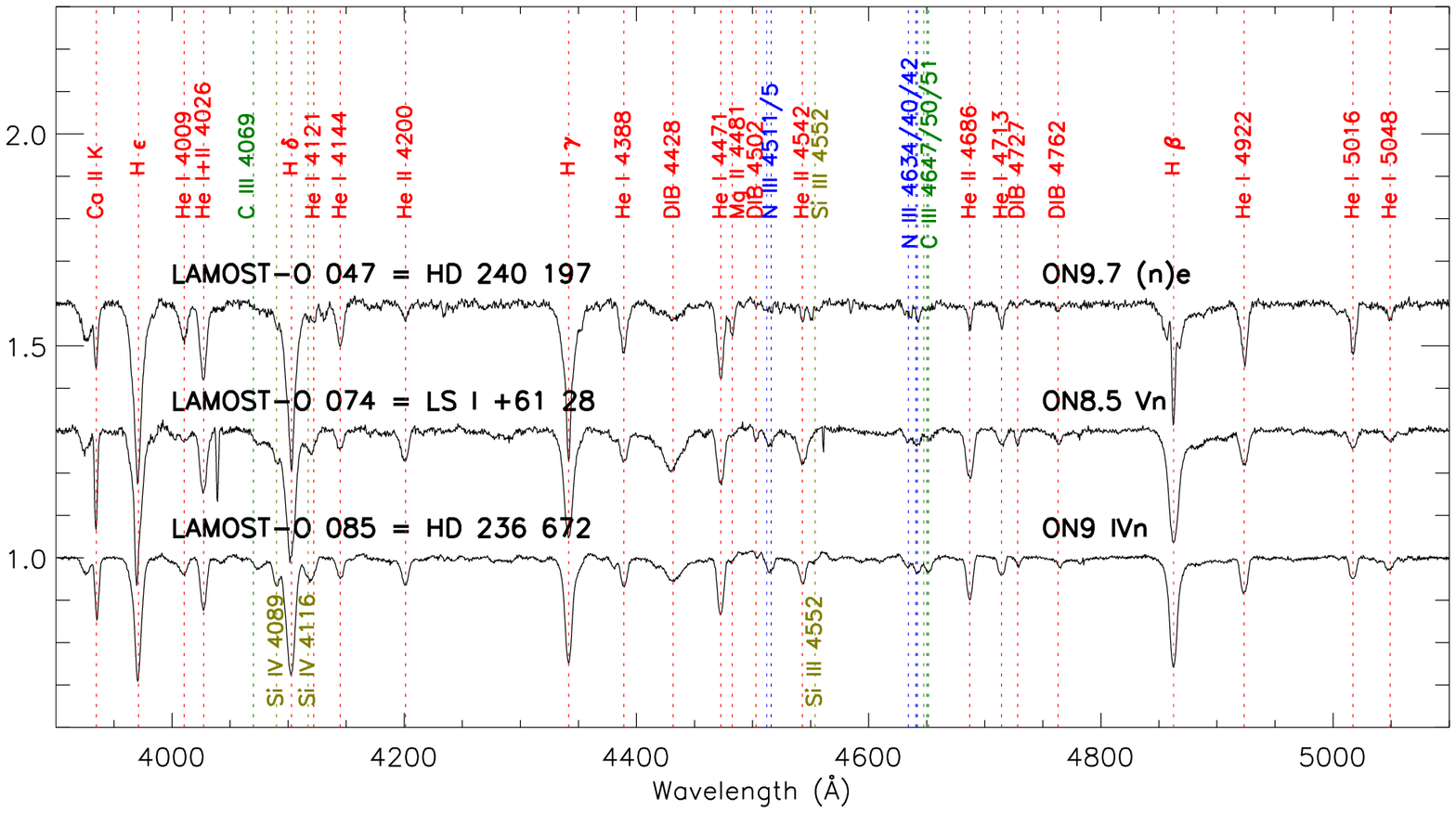}
\caption{Spectra of ON stars: LAMOST-O 047, 074, and 085.}
\label{fig:on}
\end{figure*}

The O-type stars with N\,III $\lambda$4641/42 absorption stronger than C\,III $\lambda$4650/51 absorption are classified as ON stars. Their enriched N may be produced by the nucleosynthesis , and transported  to the surface by the rotationally induced mixing \citep{li20a}. LAMOST ON stars are shown in Figure \ref{fig:on}.

LAMOST-O 074 = LS I +61 28 and LAMOST-O 085 = HD 236\,672 had been discussed in \citet{li20a}, while LAMOST-O 047 = HD 240\,197 was presented in Subsection \ref{subsec:oe}. 

\subsection{Double-lined Spectroscopic Binaries (SB2)} \label{subsec:sb2}
\begin{figure*}
\center
\includegraphics[angle=0, scale=0.8]{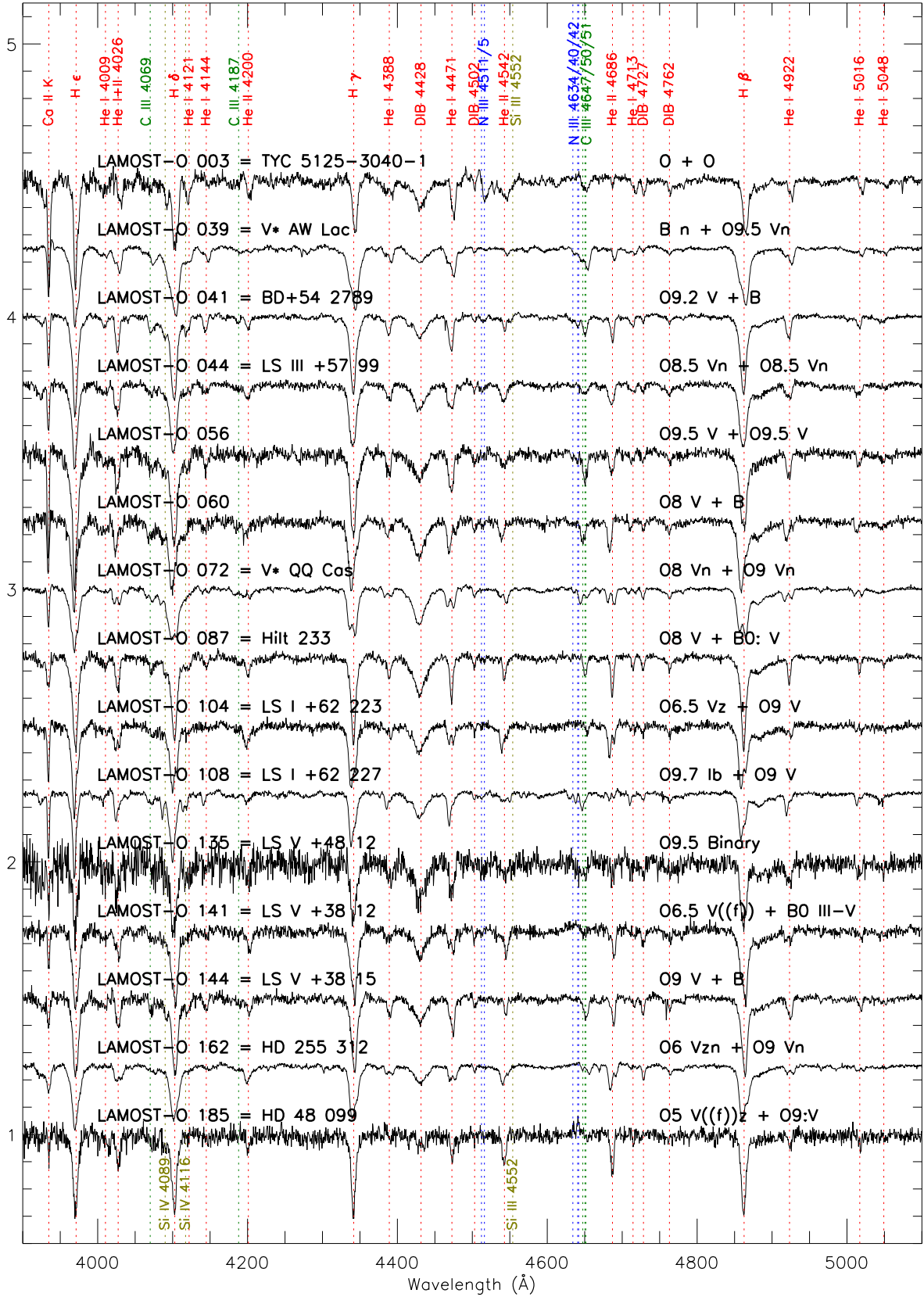}
\caption{Spectra of SB2 stars ( LAMOST-O 003, 007, 039, 041, 044,  056, 059, 060, 072, 087, 104, 108, 135, 141, 144, 162, and 185).}
\label{fig:sb2}
\end{figure*}

LAMOST SB2 stars are shown in Figure \ref{fig:sb2}. Because of the low resolution and/or quality of LAMOST spectra, the spectral types of the primary and secondary in each binary may be not correctly  determined.

\emph{LAMOST-O 003 = TYC 5125-3040-1}. In Simbad, there is no spectral information about this star. Thus, it may be a new SB2. The period is 4.3326 days \citep{sha14}.
 
\emph{LAMOST-O 039 = V* AW Lac}. In Simbad, it is a $\beta$ Lyrae type eclipsing system, with a period of 1.14285456 days \citep{jiang83}. In \citet{skiff14}, its spectral type is OB or B type. Its LAMOST spectrum is an SB2.

 \emph{LAMOST-O 041 = BD+54 2789}. In Simbad, it is a double or multiple system. In \citet{skiff14}, its spectral type is OB. There are too many stars crowded around BD+54 2789, so the LAMOST spectrum may be contaminated. 

\emph{LAMOST-O 044 = LS III +57 99}. In \citet{skiff14}, its spectral type is OB-. In \citet{sha14}, it is a binary with a period of 1.8432649 days. 
 
\emph{LAMOST-O 056}. In \citet{skiff14}, its spectral type is B. Its LAMOST spectrum is an SB2.

 \emph{LAMOST-O 060}. In \citet{skiff14}, its spectral type is O8.5 Ve \citep{chi84} or O9 V \citep{rus07}. It is also suggested to be an eclipsing binary with a period of 1.1409 days \citep{sha14}.
  
 \emph{LAMOST-O 072 = V* QQ Cas}. In \citet{skiff14}, its spectral type is OB or B type. In Simbad, it is a $\beta$ Lyrae type eclipsing system, with a period of 2.142030 days \citep{kuk57}.  
 
\emph{LAMOST-O 087 = Hilt 233}. This star has been resolved by GOSSS as an SB2 \citep{mai16}, which is adopted in this paper.

\emph{LAMOST-O 104 = LS I +62 223}. In \citet{skiff14}, its spectral type is OB+. The LAMOST spectrum is an O binary.
 
\emph{LAMOST-O 108 = LS I +62 227}. In \citet{skiff14}, its spectral type is OB, OB$+$e, or B2. The LAMOST spectrum is an O binary.
 
 \emph{LAMOST-O 109}. This star is an SB2, which will be discussed in a future paper and not shown in Figure \ref{fig:sb2}.
 
\emph{LAMOST-O 135 = LS V +48 12}.  In \citet{roman19}, its spectral type is O8.5 III(f). But there is no N III $\lambda$4634-40-42 emission. Its LAMOST median-resolution spectra show it is an SB2 in Figure \ref{fig:oe_ha}. If two stars in the binary have a similar mass, then their spectral type should be around O9.5. 

\emph{LAMOST-O 141 = LS V +38 12}. This star is assigned a spectral type of O7V((f))+B0III-V by \citet{mai16}. In Figure \ref{fig:sb2}, the He II $\lambda$4542 is slightly stronger than He I $\lambda$4471 for the primary, so it should be O6.5 V((f))+B0 III-V.

\emph{LAMOST-O 144 = LS V +38 15}. In \citet{roman19}, its spectral type is O8.5 IV(n), but the asymmetries of He I $\lambda$4388, 4471, 4922, and He II $\lambda$4686 suggest it is an SB2.

\emph{LAMOST-O 162 = HD 255\,312}.\textbf{ It was observed by the Kepler telescope in the K2 mission with a quasi-period of 0.734701 days \citep{arm15}. However, \citet{sha14} suggested it is an eclipsing binary with a period of 1.46976 days.}

 \emph{LAMOST-O 185 = HD 48\,099}. The LAMOST spectrum is 15.73'' away from HD 48\,099. This star is also in GOSSS \citep{sota14}. The signal-to-noise ratio of the LAMOST spectrum is low, but the spectra of LAMOST and GOSSS seem well consistent with each other. Thus, the GOSSS spectral type O5 V((f))z + O9: V is adopted.

\subsection{Single-lined Spectroscopic Binaries (SB1)} \label{subsec:sb1}
\begin{figure*}
\center
\includegraphics[angle=0, scale=0.8]{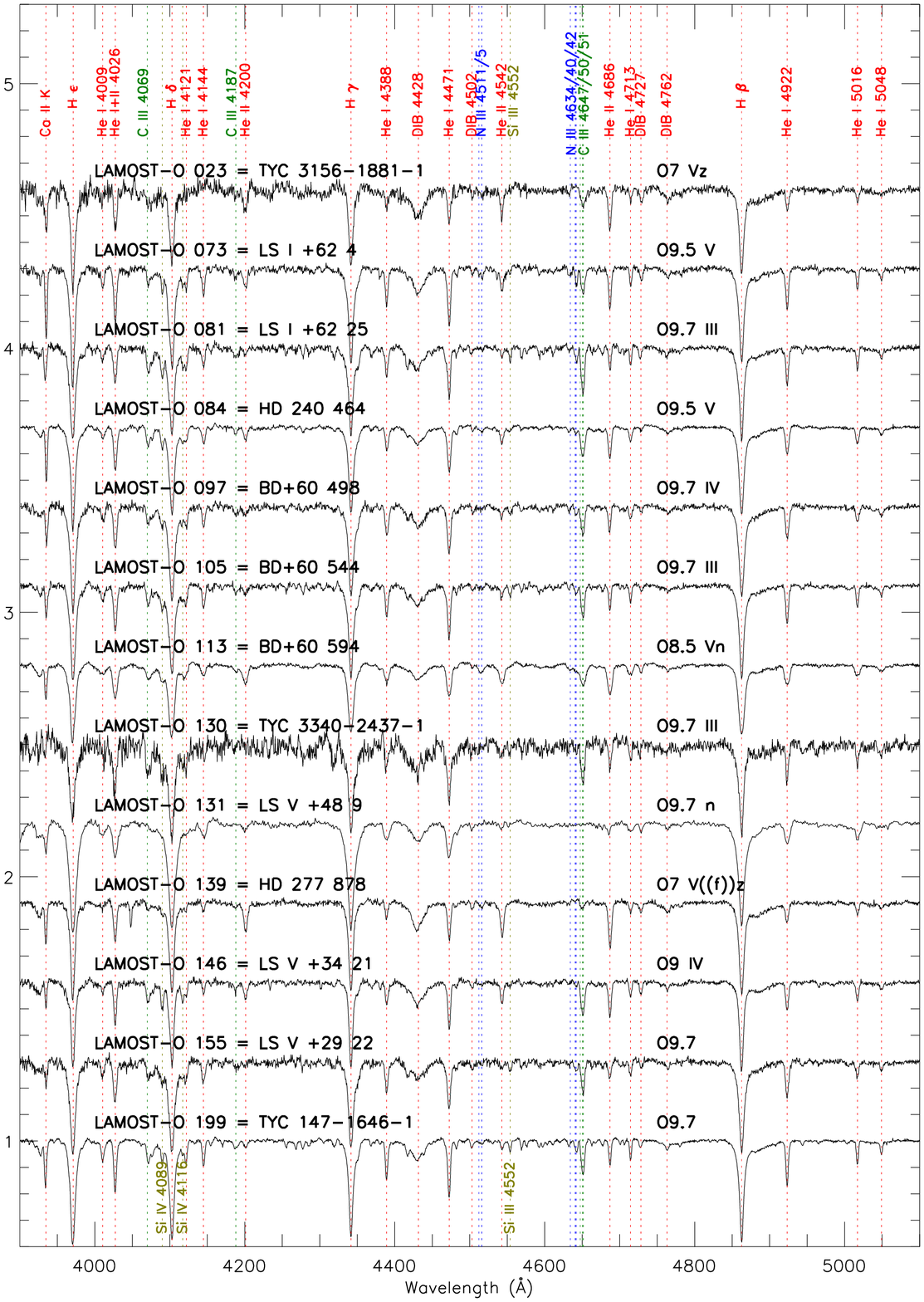}
\caption{Spectra of SB1 stars ( LAMOST-O 023, 073, 081, 084, 097, 105, 113, 130, 131, 139, 146, 155, and 199).}
\label{fig:sb1}
\end{figure*}

SB1 stars are shown in Figure \ref{fig:sb1}. Each star was observed by LAMOST for at least two times and shows an obvious shift in wavelength between its spectra from different epochs. LAMOST-O 023, 073, and 146 are determined to be SB1 stars from their medium-resolution spectra from different epochs, while others are determined from low-resolution spectra.

\emph{LAMOST-O 023 = TYC 3156-1881-1}. In \citet{ber18}, its spectral type is O8 IIIz. The LAMOST spectrum is O7 Vz. According to its medium-resolution spectra, it is an SB1, with $\Delta V \sim 75$ km\,s$^{-1}$.

\emph{LAMOST-O 073 = LS I +62 4}. In \citet{skiff14}, its spectral type is OB-. Its LAMOST spectral type is O9.5 V. According to its medium-resolution spectra, it is an SB1, with $\Delta V \sim 43$ km\,s$^{-1}$.

\emph{LAMOST-O 081 = LS I +62 25}. In \citet{skiff14}, its spectral type is OB. Its LAMOST spectral type is O9.7 III, with $\Delta V \sim 168$ km\,s$^{-1}$. 

\emph{LAMOST-O 084 = HD 240\,464}. \citet{martin72} classified this star as O9 V. Its LAMOST spectral type is O9.5 V, with $\Delta V \sim 90$ km\,s$^{-1}$. 

\emph{LAMOST-O 097 = BD+60 498}. This star has a spectral type of O9.7 II-III in \citet{sota11}, but by comparing spectra from LAMOST and GOSSS, its He I $\lambda$4713 in the LAMOST spectrum is slightly weaker than that in the GOSSS spectrum. Thus, it is classified as O9.7 IV. Besides, there is an obvious wavelength shift between two spectra ($\Delta V \sim 96$ km\,s$^{-1}$), which suggests it is an SB1.  

\emph{LAMOST-O 105 = BD+60 544}. In Simbad, this star is a double or multiple system, but without reference. In \citet{ryd78}, its spectral type is OB:. Its LAMOST spectral type is O9.7 III, with $\Delta V \sim 22$ km\,s$^{-1}$. 

\emph{LAMOST-O 113 = BD+60 594}. In \citet{ryd78}, its spectral type is O9 V. Its LAMOST spectral type is O8.5 Vn, with $\Delta V \sim 60$ km\,s$^{-1}$. 

\emph{LAMOST-O 130 = TYC 3340-2437-1}. In \citet{liu19}, its spectral type is B. Its LAMOST spectral type is O9.7 III, with $\Delta V \sim 60$ km\,s$^{-1}$. 

 \emph{LAMOST-O 131 = LS V +48 9}. In \citet{roman19}, its spectral type is O9.7 IVn. Its weak He II $\lambda$4686 indicates a high luminosity, but its weak Si IV $\lambda$4089 and 4116 imply a low one. Thus, its luminosity is unclear.  As a result, it is assigned O9.7 n. Its He II line profiles show obvious variabilities or shifts, which imply it is an SB1.

\emph{LAMOST-O 139 = HD 277\,878}. In \citet{roman19}, its spectral type is also O7 V((f))z. Moreover, there is a $\Delta V \sim 35$ km\,s$^{-1}$ between LAMOST spectra. 

\emph{LAMOST-O 146 = LS V +34 21}.  In \citet{roman19}, its spectral type is also O9 IV. According to its medium-resolution spectra, it is an SB1, with $\Delta V \sim 18$ km\,s$^{-1}$.

\emph{LAMOST-O 155 = LS V +29 22}. In \citet{skiff14}, its spectral type is OB. Its LAMOST spectral type is O9.7, with an unclear luminosity. LAMOST spectra also have a $\Delta V \sim 52$ km\,s$^{-1}$. 

\emph{LAMOST-O 199 = TYC 147-1646-1}. In \citet{mof79}, its spectral type is B0 III. For a similar reason to LAMOST-O 131, its luminosity is unclear. Though $\Delta V \sim 18$ km\,s$^{-1}$ is slightly low, it should be an SB1 from direct comparison of two high signal-to-noise spectra.

\subsection{Vz} \label{subsec:vz}

\begin{figure*}
\center
\includegraphics[angle=0, scale=0.8]{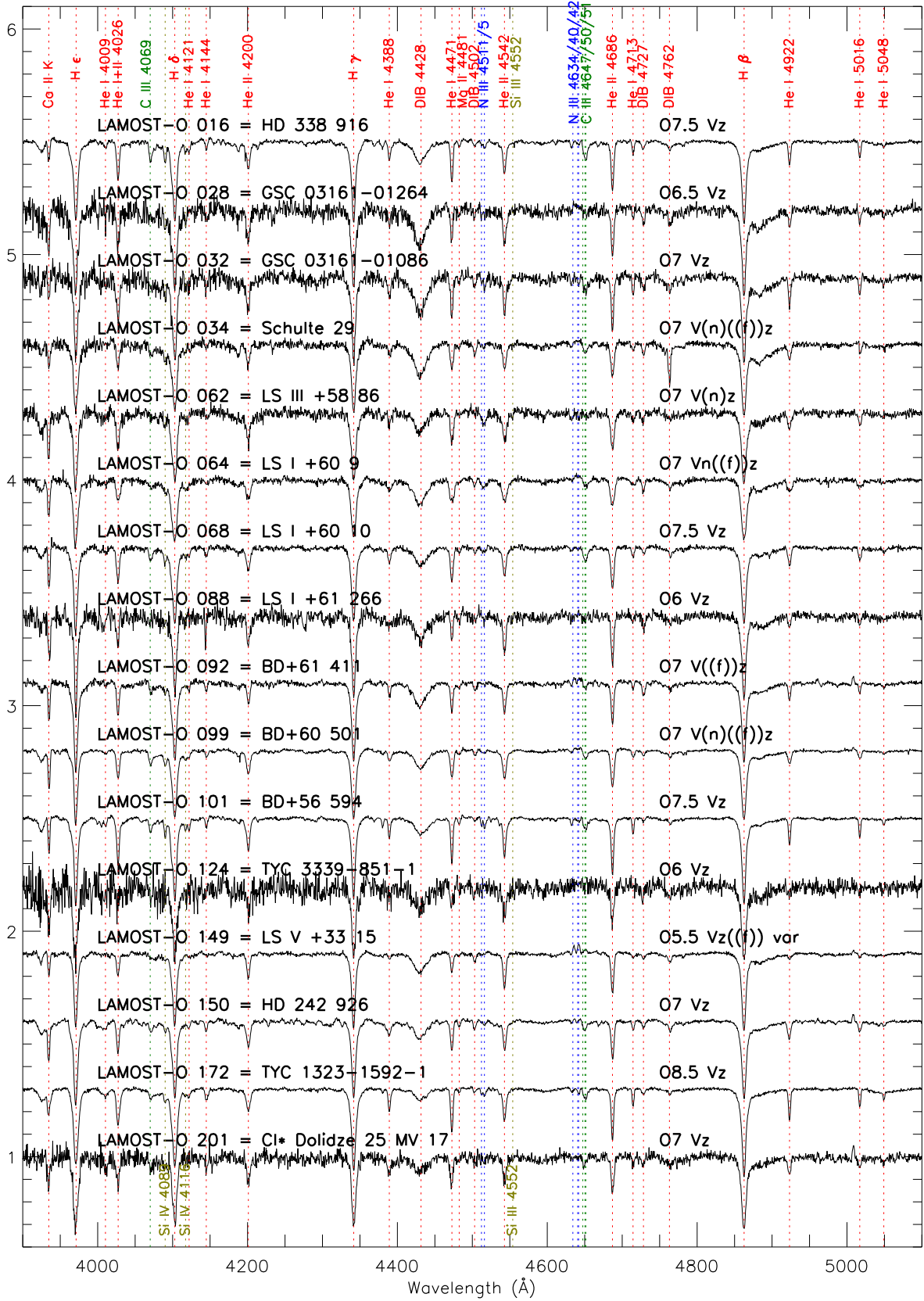}
\caption{Spectra of Vz stars: LAMOST-O 016, 028, 032, 034, 062, 064, 068, 088, 092, 099, 101, 124, 149, 150, 172, and 201.}
\label{fig:vz}
\end{figure*}

OVz stars are on or very near the zero-age main-sequence O \citep{wal09}. Thus, they are among the youngest massive stars \citep{arias16}. In Figure \ref{fig:vz}, 14 OVz stars in LAMOST data are shown. LMOST-O 023 and 139 are SB1 and have been presented in Subsection \ref{subsec:sb1}.

\emph{LAMOST-O 016 = HD 338\,916}. This star is also in GOSSS data \citep{mai16}, with a spectral type of O7.5 Vz. There is no radial velocity variation between LAMOST and GOSSS spectra, but He I $\lambda$4471 and He II $\lambda$4542 in the LAMOST spectrum are a little stronger.

\emph{LAMOST-O 028 = GSC 03161-01264} and \emph{LAMOST-O 032 = GSC 03161-01086}. These two stars are O6.5 V and O7 V, respectively, in \citet{com12}. However, their He II $\lambda$4686 in LAMOST spectra are strong enough to be Oz. Thus, they are O6.5 Vz and O7 Vz, respectively.

\emph{LAMOST-O 034 = Schulte 29}.  Its GOSSS spectral type is O7.5 V(n)((f))z \citep{mai16}. But its He I $\lambda$4471 in LAMOST spectrum is a little weaker than that in the GOSSS spectrum, thus it is O7 V(n)((f))z. It is an SB1 \citep{kob14}, but its LAMOST and GOSSS spectra do not show obvious radial velocity variation. The reason may be that the semiamplitude of the radial velocity curve of the primary, 17.5 km\,s$^{-1}$, is too low to be detected.

\emph{LAMOST-O 062 = LS III +58 86}. This star was categorized as B2 in \citet{bro53}. Its LAMOST spectral type is O7 V(n)z.

\emph{LAMOST-O 064 = LS I +60 9}. It is a B-type star, without reference in Simbad. Its LAMOST spectral type is O7 Vn((f))z.

\emph{LAMOST-O 068 = LS I +60 10}. In \citet{bro53}, it is B2, but its LAMOST spectral type is O7.5 Vz.

\emph{LAMOST-O 088 = LS I +61 266}. It is a B-type star, without a reference in Simbad. LAMOST spectral type is O6 Vz.

\emph{LAMOST-O 092 = BD+61 411}. The spectral type is O6.5 V((f))z in \citet{mai16}, but the ratio of He II $\lambda$4542/He I $\lambda$4471 in the LAMOST spectrum is slightly smaller than that in the GOSSS spectrum, thus the LAMOST spectral classification is O7 V((f))z. Moreover, there is a $\Delta V \sim 55$ km\,s$^{-1}$ between LAMOST and GOSSS spectra. Thus, it is an SB1.

\emph{LAMOST-O 099 = BD+60 501}. There is a $\Delta V \sim 60$ km\,s$^{-1}$ between LAMOST and GOSSS spectra \citep{sota11}. Thus, it is an SB1. 

\emph{LAMOST-O 101 = BD+56 594}. In \citet{ryd78}, its spectral type is OB. Its LAMOST spectral type is O7.5 Vz.

\emph{LAMOST-O 124 = TYC 3339-851-1}. This star is O5 V((f)) in \citet{roman19}, but the ratio of He II $\lambda$4542/He I $\lambda$4471 is not that large. Moreover, its He II $\lambda$4686 is stronger than He II $\lambda$4542 and He I $\lambda$4471, but N III $\lambda$4634-40-42 cannot be undoubtedly identified. Thus, it is assigned as O6 Vz.

\emph{LAMOST-O 149 = LS V +33 15}.  This star is also in \citet{mai16} with a spectral type of O7 V(n)z. But both He II $\lambda$4200/He I + II $\lambda$4026 and He II $\lambda$4542/He I $\lambda$4471 are higher in LAMOST spectrum, which suggests its spectral type should be O5.5.  Surprisingly, the strong absorption line C III $\lambda$4647/50/51 in the GOSSS spectrum becomes emission in the LAMOST spectrum, while N III $\lambda$4636/40/42 is also in emission in the LAMOST spectrum. There is a wavelength shift between the GOSSS and LAMOST spectra ($\Delta V \sim 20$ km\,s$^{-1}$), and the H$\alpha$ line also shows a variation between two LAMOST spectra. Thus, it should be an SB1, which may be responsible for the spectral variation.

\emph{LAMOST-O 150 = HD 242\,926}. It is a spectral classification standard star of O7 Vz \citep{mai16}. But there are still line variabilities between LAMOST and GOSSS spectra. Specially, He II $\lambda$4686 in the LAMOST spectrum is slightly weaker than that in the GOSSS spectrum.

\emph{LAMOST-O 172 = TYC 1323-1592-1}. It is O7.5 V((f)) in \citet{roman19}, but it should be O8.5 Vz.

\emph{LAMOST-O 201 =  Cl* Dolidze 25 MV 17}. In \citet{neg15}, its spectral type is also O7 Vz.

\subsection{Normal stars} \label{subsec:normal}

\begin{figure*}
\center
\includegraphics[angle=0, scale=0.8]{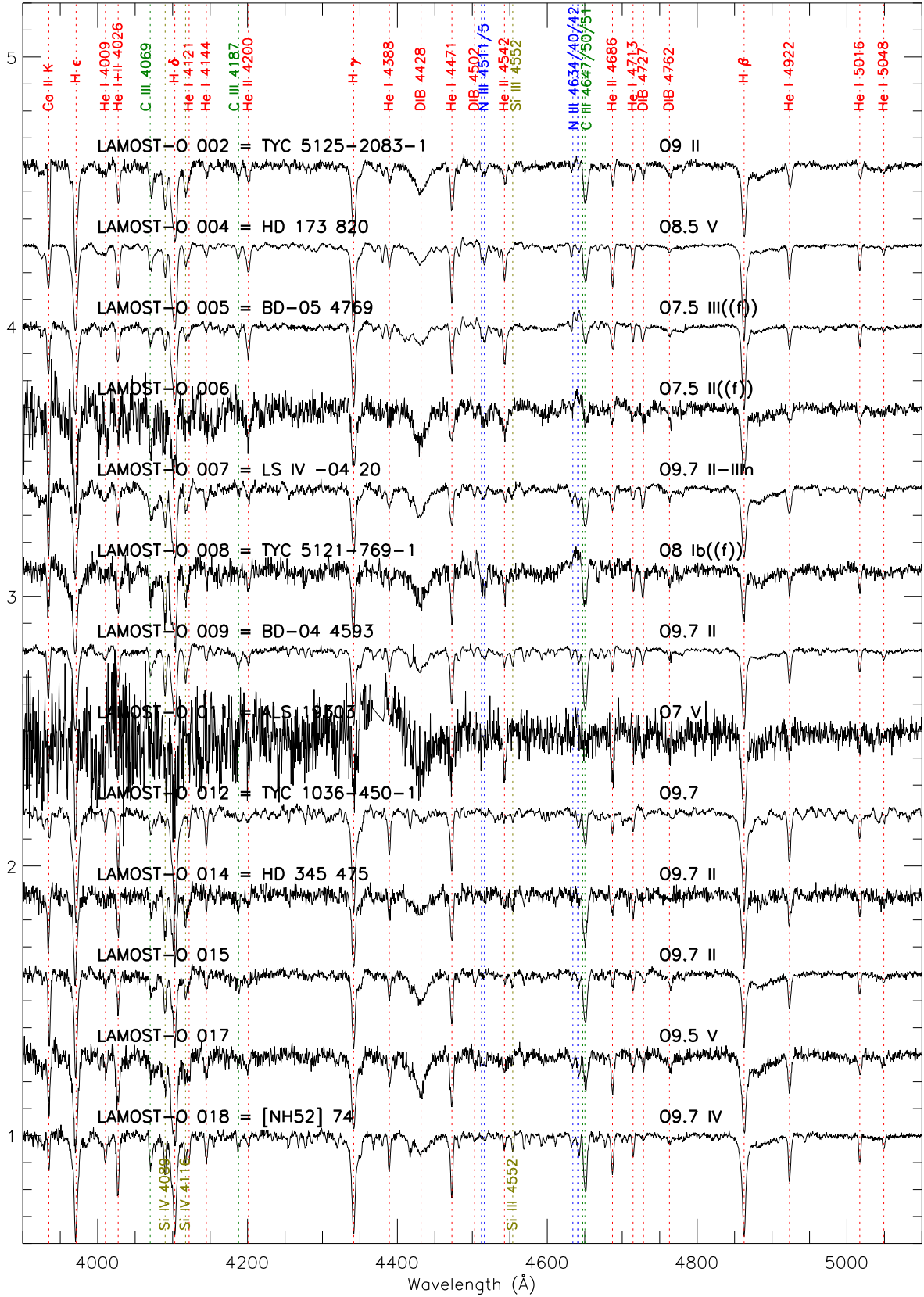}
\caption{Spectra of LAMOST-O 002, 004, 005, 006,007,  008, 009, 011, 012, 014, 015, 017, and 018.}
\label{fig:od1}
\end{figure*}

The following 13 stars are shown in Figure \ref{fig:od1}.

In Simbad, there is no spectral information about  \emph{LAMOST-O 006}, \emph{LAMOST-O 008 = TYC 5121-769-1}, \emph{LAMOST-O 015},  and \emph{LAMOST-O 017}. Thus, these four stars may be new O-type stars. 

\emph{LAMOST-O 002 = TYC 5125-2083-1}.  Its spectral type is B0 in \citet{ros63}. Its LAMOST spectrum is O9 II. 

\emph{LAMOST-O 004 = HD 173\,820}.  Its spectral type is B0.5 Ia/ab in \citet{houk99}. Its LAMOST spectrum is O8.5 V. 

\emph{LAMOST-O 005 = BD-05 4769}.  Its spectral type is O7 III in \citet{bis82}. Its LAMOST spectrum is O7.5 III((f)). 

\emph{LAMOST-O 007 = LS IV -04 20}. In \citet{skiff14}, its spectral type is OB or B2 Ib. Its LAMOST spectrum is an O 9.7 II-IIIn.

\emph{LAMOST-O 009 = BD-04 4593}. Its spectral type is O9.5 Ia in \citet{vij93}. Its LAMOST spectrum is O9.7 II. 

\emph{LAMOST-O 011 = ALS 19303}. Its spectral type is B1 II in \citet{for89}. Its LAMOST spectrum is roughly O7 V. 

\emph{LAMOST-O 012 = TYC 1036-450-1}. This star has unclear luminosity \citep{mai16}. However, there is an obvious wavelength shift between the LAMOST and GOSSS spectra ($\Delta V \sim 78$ km\,s$^{-1}$). Thus, it is an SB1. 

\emph{LAMOST-O 014 = HD 345 475}. Its spectral type is B0 in \citet{pop50}. Its LAMOST spectrum is roughly O9.7 II. 

\emph{LAMOST-O 018 = [NH52] 74}. Its spectral type is also O9.7 IV in \citet{roman19}. 

\begin{figure*}
\center
\includegraphics[angle=0, scale=0.8]{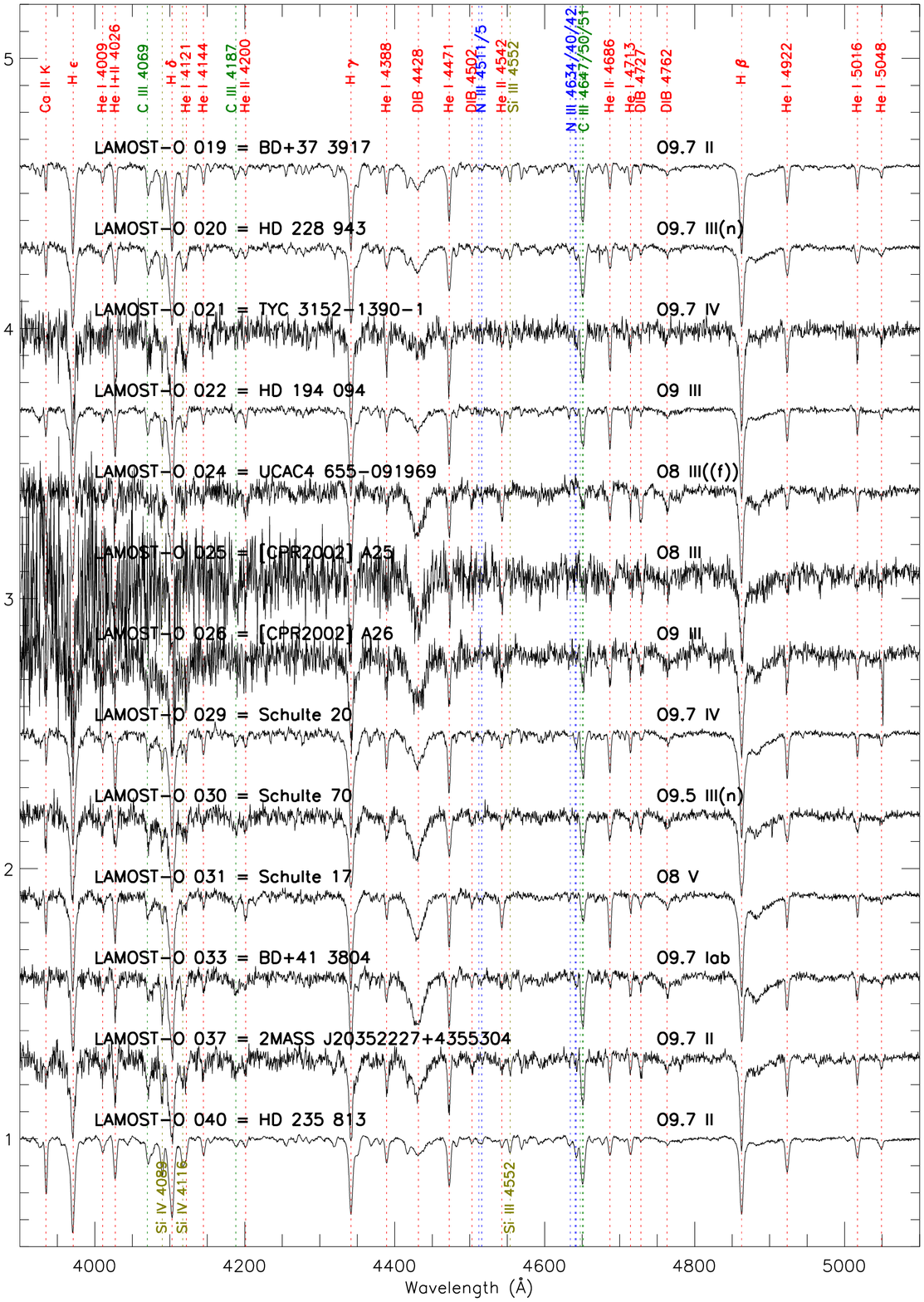}
\caption{Spectra of LAMOST-O 019, 020, 021, 022, 024, 025, 026, 029, 030, 031, 033, 037, and 040}
\label{fig:od2}
\end{figure*}

%----- 2 -----------------------------------------------------
The following 13 stars are shown in Figure \ref{fig:od2}.

\emph{LAMOST-O 019 = BD+37 3917}.  Its spectral type is OB in \citet{nas52}. Its LAMOST spectrum is roughly O9.7 II.

\emph{LAMOST-O 020 = HD 228\,943}. Its spectral type is O9.7 III in \citet{math89}. Its LAMOST spectrum is O9.7 III(n). Its radial velocity in the LAMOST spectrum is more than 80 km\,s$^{-1}$ lower than that given by \citet{huang06}, which suggests this star may be an SB1.

\emph{LAMOST-O 021 = TYC 3152-1390-1}. This star has no spectral information in Simbad.  

\emph{LAMOST-O 022 = HD 194\,094}. Its spectral type is O8.5 III in \citet{mahy13}. Its LAMOST spectrum is O9 III. Its radial velocities given by \citet{evans67}, \citet{mahy13} and the LAMOST spectrum  agrees well with each other within 10 km\,s$^{-1}$. Thus, it is likely a single star.

\emph{LAMOST-O 024 = UCAC4 655-091969}. Its spectral type is also O8 III((f)) in \citet{roman19}. 

\emph{LAMOST-O 025 = [CPR2002] A25}. Its spectrum is too noisy to be used for classification. I adopt the spectral type O8 III given by \citet{com12}. 

\emph{LAMOST-O 026 = [CPR2002] A26}. Its spectral type is O9 V in \citet{com12}. But the ratio of He II $\lambda$4686/He I $\lambda$4713 in the LAMOST spectrum suggests that its luminosity should be III. 
 
\emph{LAMOST-O 029 = Schulte 20}. Its spectral type is also O9 III in \citet{com12}. LAMOST spectrum is O9.7 IV. In \citet{kob14}, it is an SB1.

\emph{LAMOST-O 030 = Schulte 70}.  Its spectral type is O9.5 IV(n) in \citet{mai16}, but its LAMOST spectrum is more like O9.5 III(n), because of higher Si IV $\lambda$4089/He I $\lambda$4026. In \citet{kob14}, it is an SB1, with a period of 245.1 days. 

\emph{LAMOST-O 031 = Schulte 17}. This star is also in \citet{mai16}. There is a wavelength shift between GOSSS and LAMOST spectra ($\Delta V \sim 36$ km s$^{-1}$). In \citet{kob14}, it is an SB1. 

\emph{LAMOST-O 033 = BD+41 3804}. This star is also in \citet{mai16}. 

\emph{LAMOST-O 037 = 2MASS J20352227+4355304}. This star has been given in \citet{roman19}. 
 
\emph{LAMOST-O 040 = HD 235\,813}.  Its spectral type is B0 III in \citet{suad12}. Its LAMOST spectrum is O9.7 II.

\begin{figure*}
\center
\includegraphics[angle=0, scale=0.8]{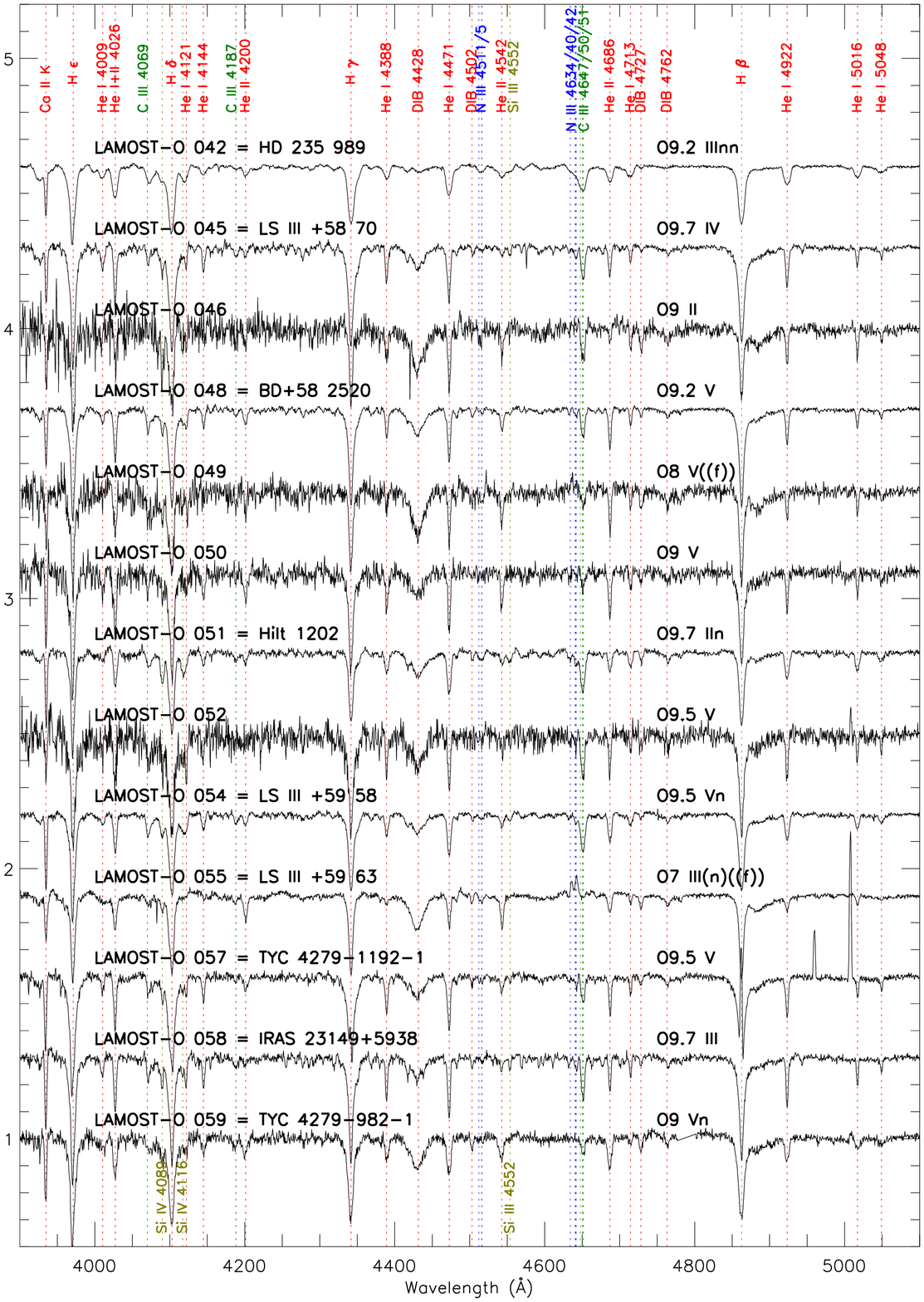}
\caption{Spectra of LAMOST-O 042, 045, 046, 048, 049, 050, 051, 052, 054, 055, 057, 058, and 059.}
\label{fig:od3}
\end{figure*}

% ----- 3 ------
 The following 13 stars are shown in Figure \ref{fig:od3}.
 
In Simbad, there is no spectral information about  \emph{LAMOST-O 046}, \emph{LAMOST-O 049}, \emph{LAMOST-O 050}, \emph{LAMOST-O 052}, and \emph{LAMOST-O 059}. Thus, these four stars may be new O-type stars. 
 
 \emph{LAMOST-O 042 = HD 235\,989}. In \citet{skiff14}, its spectral type is B or OB. Its LAMOST spectrum is O9.2 IIInn.

\emph{LAMOST-O 045 = LS III +58 70}. Its spectral type is B0 V in \citet{neg03}. LAMOST spectrum is O9.7 IV. 

\emph{LAMOST-O 048 = BD+58 2520}.  Its spectral type is B2 in \citet{bro53}. Its LAMOST spectrum is O9.2 V. 

\emph{LAMOST-O 051 = Hilt 1202}. In \citet{skiff14}, its spectral type is B or OB. Its LAMOST spectrum is O9.7 IIn. 

\emph{LAMOST-O 054 = LS III +59 58}. In \citet{skiff14}, its spectral type is B or OB. Its LAMOST spectrum is O9.5 Vn. 

\emph{LAMOST-O 055 = LS III +59 63}. In \citet{skiff14}, its spectral type is B, OB, or even A8. Its LAMOST spectrum is O7 III(n)(f). 

\emph{LAMOST-O 057 = TYC 4279-1192-1}. In \citet{chi84}, its spectral type is also O9.5 V.

\emph{LAMOST-O 058 = IRAS 23149+5938}. In \citet{skiff14}, its spectral type is F. Its LAMOST spectrum is O9.7 III. 

\begin{figure*}
\center
\includegraphics[angle=0, scale=0.8]{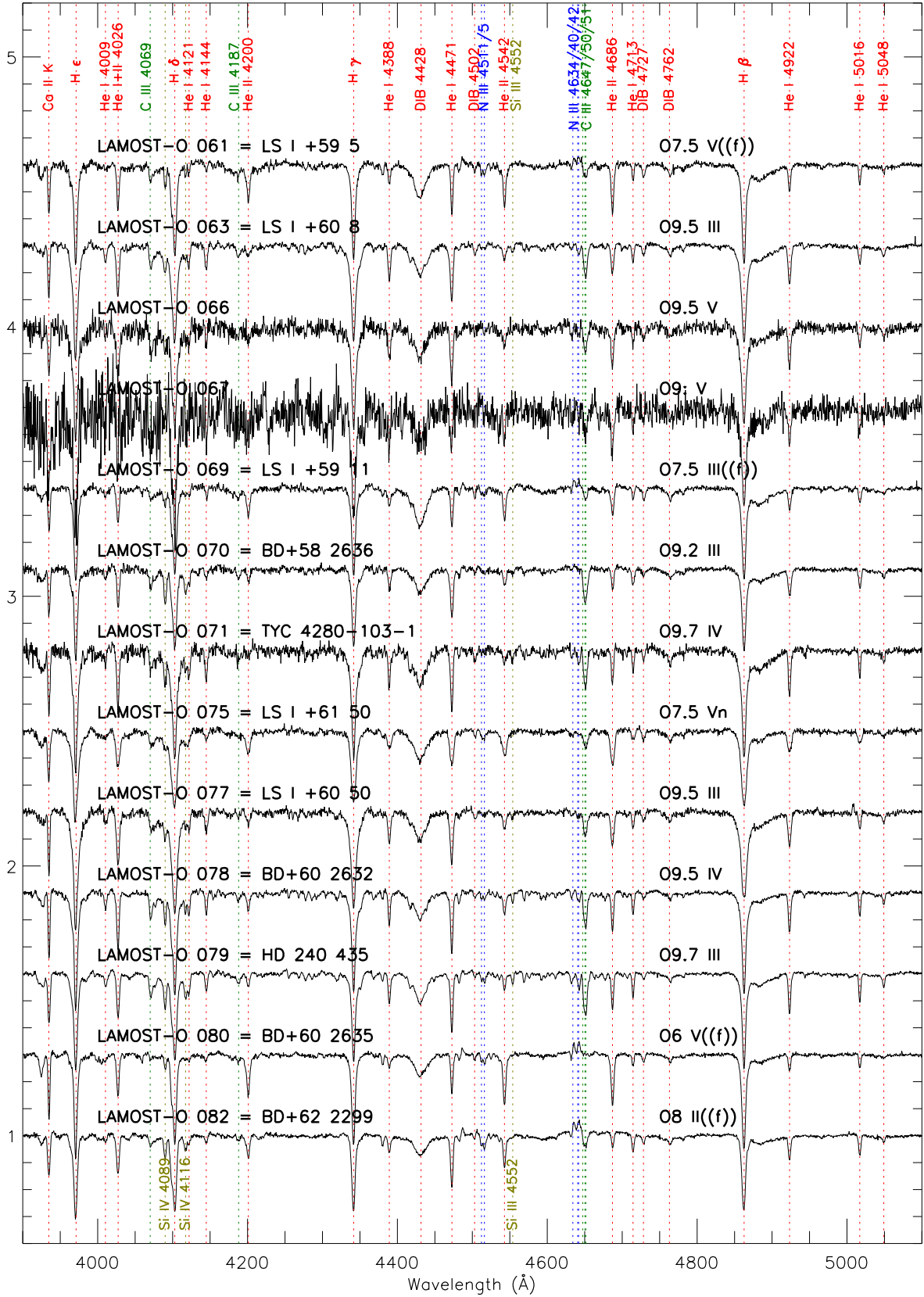}
\caption{Spectra of LAMOST-O 061, 063, 066, 067, 069, 070, 071, 075, 077, 078, 079, 080, and 082.}
\label{fig:od4}
\end{figure*}

%-------4--------
The following 13 stars are shown in Figure \ref{fig:od4}.

\emph{LAMOST-O 061 = LS I +59 5}. In \citet{skiff14}, its spectral type is B0 or OB. Its LAMOST spectrum is O7.5 V((f)).

\emph{LAMOST-O 063 = LS I +60 8}. In \citet{curz74}, its spectral type is O9.5 V. Its LAMOST spectrum is O9.5 III. 

\emph{LAMOST-O 066}. In \citet{rus07}, its spectral type is O9 V. Its LAMOST spectrum is O9.5 V. 

\emph{LAMOST-O 067}. In \citet{rus07}, its spectral type is O8 V. The ratio of signal to noise of the LAMOST spectrum is very low, but it seems that the ratio of He I $\lambda$4542/He I $\lambda$4388 $\sim$ 1, so it is assigned O9: V. 

\emph{LAMOST-O 069 = LS I +59 11}. In \citet{skiff14}, its spectral type is B or OB. Its LAMOST spectrum is O7.5 III((f)). 

\emph{LAMOST-O 070 = BD+58 2636}. In \citet{skiff14}, its spectral type is B or OB. Its LAMOST spectrum is O9.2 III. 

\emph{LAMOST-O 071 = TYC 4280-103-1}. This star has no spectral information in Simbad.  

\emph{LAMOST-O 075 = LS I +61 50}.  In \citet{skiff14}, its spectral type is B2 or OB. Its LAMOST spectrum is O7.5 Vn. 

\emph{LAMOST-O 077 = LS I +60 50}. In \citet{hunter90}, its spectral type is also O9.5 III.

\emph{LAMOST-O 078 = BD+60 2632}. In \citet{skiff14}, its spectral type is B or OB. Its LAMOST spectrum is O9.5 IV. 

\emph{LAMOST-O 079 = HD 240\,435}. The spectral type in \citet{reed03} is B0.5.Its  LAMOST spectrum is O9.7 III. 

\emph{LAMOST-O 080 = BD+60 2635}. This star is also in \citet{mai16}. There is a wavelength shift between the GOSSS and LAMOST spectra ($\Delta V \sim 74$ km\,s$^{-1}$). Thus, it is an SB1. 

\emph{LAMOST-O 082 = BD+62 2299}. In \citet{skiff14}, its spectral type is B or OB. Its LAMOST spectrum is O8 II((f)).

%--------------5--------

\begin{figure*}
\center
\includegraphics[angle=0, scale=0.8]{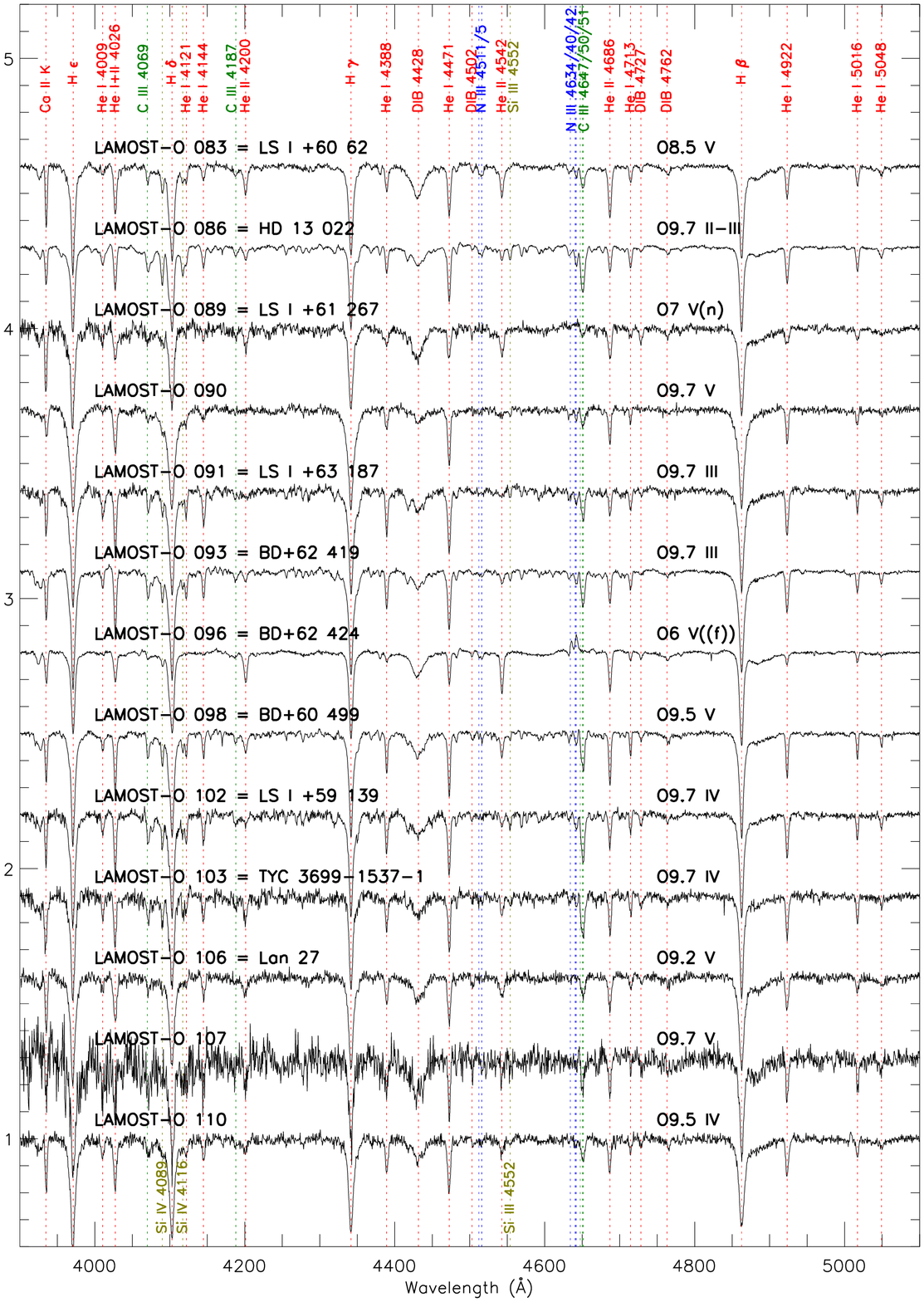}
\caption{Spectra of LAMOST-O 083, 086, 089, 090, 091, 093, 096, 098, 102, 103, 106, 107, and 110.}
\label{fig:od5}
\end{figure*}

The following 13 stars are shown in Figure \ref{fig:od5}.

\emph{LAMOST-O 083 = LS I +60 62}. In \citet{skiff14}, its spectral type is B or OB. Its LAMOST spectrum is O8.5 V.

\emph{LAMOST-O 086 = HD 13\,022}. This star is also in \citet{sota11}. There is a wavelength shift between GOSSS and LAMOST spectra ($\Delta V \sim 69$ km\,s$^{-1}$). Thus, it is an SB1.

\emph{LAMOST-O 089 = LS I +61 267}. In \citet{har59}, its spectral type is OB. Its LAMOST spectrum is O7 V(n). 

\emph{LAMOST-O 090}. This star is not in Simbad, so it may be a new O-type star. 

\emph{LAMOST-O 091 = LS I +63 187}.  In \citet{skiff14}, its spectral type is B or OB. The ratio of He I $\lambda$4686/He I $\lambda$4713 suggests its luminosity is II-III, but its Si IV$\lambda$4089 and 4116 are too weak to suggest a higher luminosity, so it is assigned as O9.7 III. 

\emph{LAMOST-O 093 = BD+62 419}. In \citet{skiff14}, its spectral type is B or OB. Its LAMOST spectrum is O9.7 III. 

\emph{LAMOST-O 096 = BD+62 424}.  This star is also in \citet{sota11}. There is a wavelength shift between the GOSSS and LAMOST spectra ($\Delta V \sim 46$ km\,s$^{-1}$). Thus, it is an SB1. The ratio of He I + II $\lambda$4200 / He II $\lambda$ 4026 $\sim$ 1, higher than that in the GOSSS spectrum. Besides, limited by the low resolution of the LAMOST spectrum, its rotation is indiscernible. Thus, its LAMOST spectrum is O6 V((f)). 

\emph{LAMOST-O 098 = BD+60 499}. This star is also in \citet{sota11}. There is a wavelength shift between the GOSSS and LAMOST spectra ($\Delta V \sim 54$ km\,s$^{-1}$). Thus, it is an SB1. 

\emph{LAMOST-O 102 = LS I +59 139}.  In \citet{har59}, its spectral type is OB. Its LAMOST spectrum is O9.7 IV. 

\emph{LAMOST-O 103 = TYC 3699-1537-1}. This star has no spectral information in Simbad.

\emph{LAMOST-O 106 = Lan 27}. In \citet{mar81}, its spectral type is late B. Its LAMOST spectrum is O9.2 V. 

\emph{LAMOST-O 107}. This star has no spectral information in Simbad.

\emph{LAMOST-O 110}. In \citet{rus07}, its spectral type is also O9.5 IV.

%--------------6--------

\begin{figure*}
\center
\includegraphics[angle=0, scale=0.8]{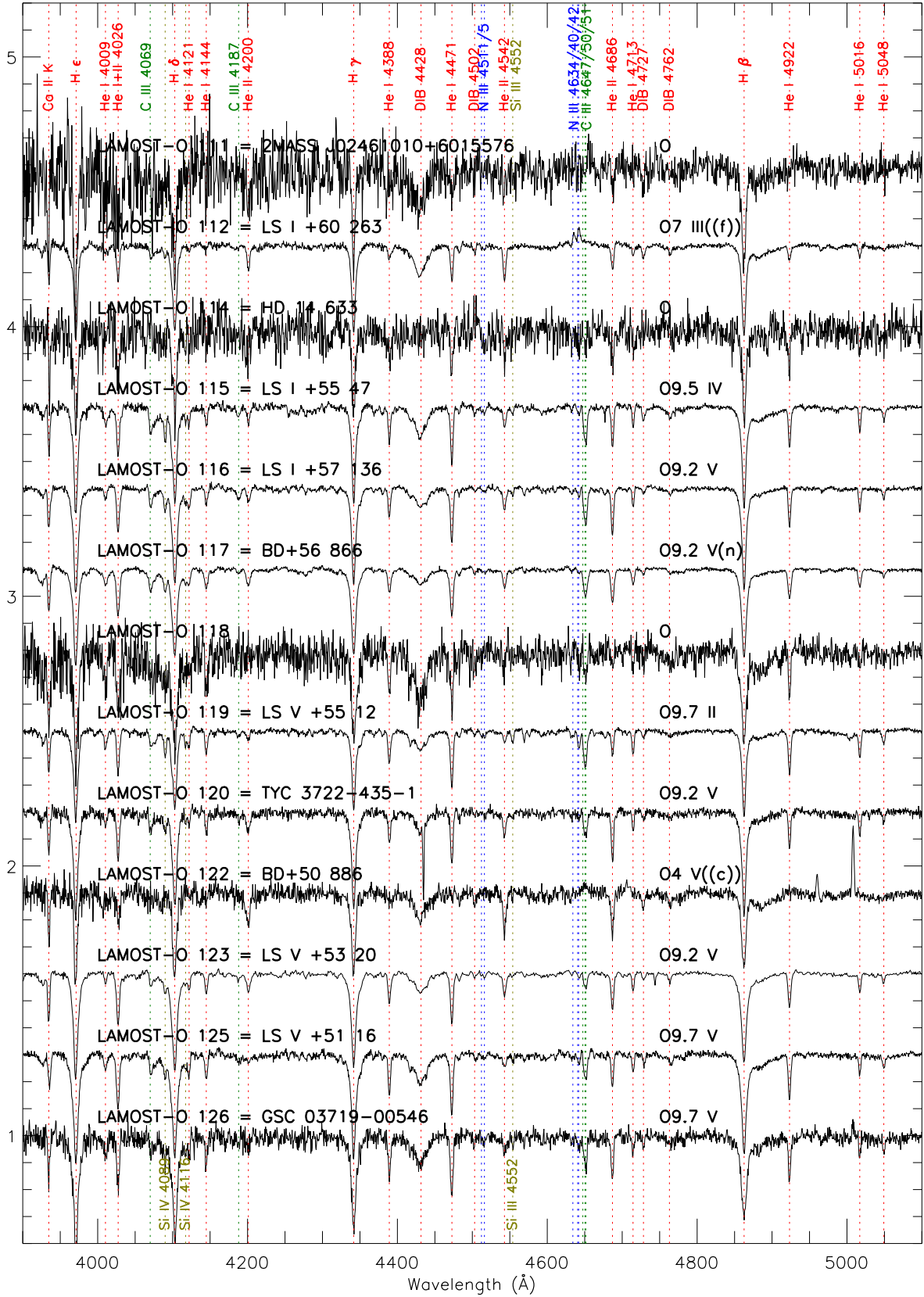}
\caption{Spectra of LAMOST-O 111, 112, 114, 115, 116, 117, 118, 119, 120, 122, 123, 125, and 126.}
\label{fig:od6}
\end{figure*}

The following 13 stars are shown in Figure \ref{fig:od6}.

\emph{LAMOST-O 111 = 2MASS J02461010+6015576}. In Simbad, it is an infrared source without spectral information. In the LAMOST spectrum, its He II $\lambda$4200 and 4686 are clearly seen, while He II $\lambda$4542 can be marginally seen, so it is a late O-type star.

\emph{LAMOST-O 112 = LS I +60 263}. In \citet{skiff14}, its spectral type is B or OB. Its LAMOST spectrum is O7 III((f)).

\emph{LAMOST-O 114}. This spectrum is 28''.34 away from HD 14\,633. By comparing with the GOSSS spectrum in \citet{sota11}, the LAMOST spectrum is well consistent with the GOSSS spectrum, but with low signal-to-noise ratio.
 
\emph{LAMOST-O 115 = LS I +55 47}. In \citet{roman19}, its spectral type is also O9.5 IV. 

\emph{LAMOST-O 116 = LS I +57 136}. In \citet{rus07}, its spectral type is O9 V. Its LAMOST spectrum is O9.2 V. 

\emph{LAMOST-O 117 = BD+56 866}.  In \citet{neg03}, its spectral type is O9 V. Its LAMOST spectrum is O9.2 V(n). 

\emph{LAMOST-O 118}.  This star has no spectral information in Simbad. Though the signal-to-noise ratio is very low, its He II $\lambda$4542 is clearly seen.

\emph{LAMOST-O 119 = LS V +55 12}.  In \citet{skiff14}, its spectral type is B0 or OB. Its LAMOST spectrum is O9.7 II.

\emph{LAMOST-O 120 = TYC 3722-435-1}. There is no spectral information for this star in Simbad.

\emph{LAMOST-O 122 = BD+50 886}.  This star is also in \citet{mai16}. There is a wavelength shift between GOSSS and LAMOST spectra ($\Delta V \sim 76$ km\,s$^{-1}$). Thus, it may be an SB1. The N III $\lambda$4634/40/42 emission cannot be identified anymore, but the C III $\lambda$4647/50/51 emission emerges. Thus, it is O4 V((c)).

\emph{LAMOST-O 123 = LS V +53 20}. In \citet{roman19}, its spectral type is O9 V. By comparing with standard stars in GOSSS, its spectral type is more like O9.2 V.

\emph{LAMOST-O 125 = LS V +51 16}. In \citet{roman19}, its spectral type is O9.2 IV(n). By comparing with standard stars in GOSSS, its spectral type is more like O9.7 V.

\emph{LAMOST-O 126 = GSC 03719-00546}. In \citet{mol18}, its spectral type is O9.5 V. Its LAMOST spectrum is O9.7 V.

%--------------7--------

\begin{figure*}
\center
\includegraphics[angle=0, scale=0.8]{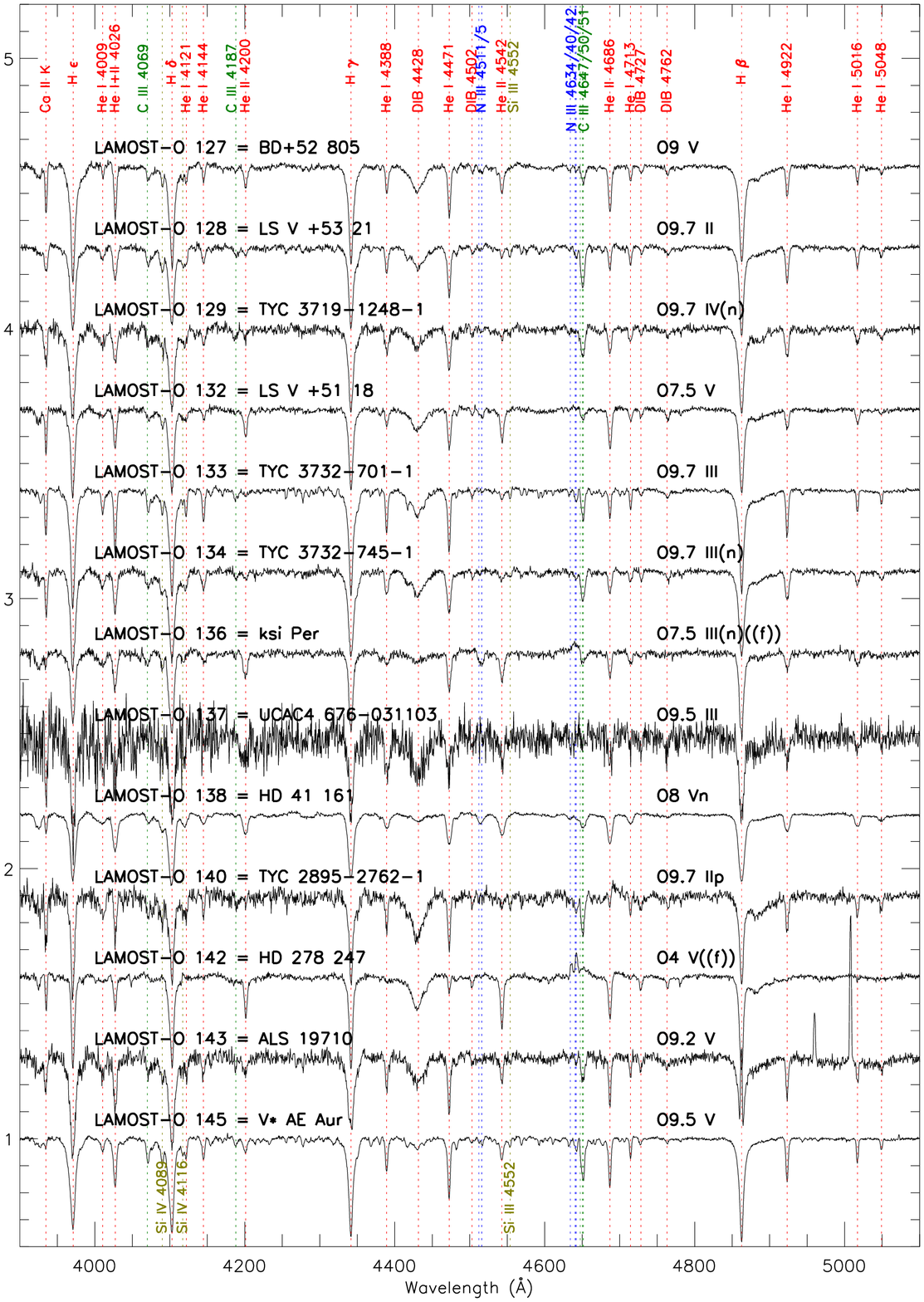}
\caption{Spectra of LAMOST-O 127, 128, 129, 132, 133, 134, 136, 137, 138, 140, 142, 143, and 145.}
\label{fig:od7}
\end{figure*}

The following 13 stars are shown in Figure \ref{fig:od7}.

\emph{LAMOST-O 127 = BD+52 805}.  This star is also in \citet{mai16} with a spectral type of O8 V(n). But both He II $\lambda$4200/He I $\lambda$4144 and He II $\lambda$4542/He I $\lambda$4388 are around 1, which suggest its spectral type should be O9. But, its rotation is indiscernible from the low resolution of the LAMOST spectrum. As a result, its spectral type should be O9 V. Moreover, it seems that there is a wavelength shift between GOSSS and LAMOST spectra ($\Delta V \sim 32$ km\,s$^{-1}$). Thus, it may be an SB1.

\emph{LAMOST-O 128 = LS V +53 21}. In \citet{har65}, its spectral type is OB+. Its LAMOST spectrum is O9.7 II.

\emph{LAMOST-O 129 = TYC 3719-1248-1}. It is an eclipsing binary with a period of 5.0254 days \citep{col18}.

\emph{LAMOST-O 132 = LS V +51 18}.  In \citet{skiff14}, its spectral type is OB. Its LAMOST spectrum is O7.5 V. 

\emph{LAMOST-O 133 = TYC 3732-701-1}. In \citet{liu19},  its spectral type is B. Its new spectral type is O9.7 III.

\emph{LAMOST-O 134 = TYC 3732-745-1}. In \citet{liu19},  its spectral type is B. Its new spectral type is O9.7 III(n).

\emph{LAMOST-O 136 = $\xi$ Per}. This spectrum is 28''.65 away from $\xi$ Per, but by comparing with its GOSSS spectrum in \citet{sota11}, its LAMOST spectrum is well consistent with its GOSSS spectrum, but with a $\Delta V \sim 42$ km\,s$^{-1}$.

\emph{LAMOST-O 137 = UCAC4 676-031103}.  In \citet{roman19}, its spectral type is O8 V((f)). But its He II $\lambda$4542 is obviously weaker than He I $\lambda$4388, which suggests its spectral type is more like O9.5, while its luminosity seems more like III based on He II $\lambda$4686/He I $\lambda$4713. As a result, it is classified as O9.5 III.
 	
\emph{LAMOST-O 138 = HD 41 161}. This spectrum is 7.48'' away from HD 41\,161, but its LAMOST spectrum is consistent well with GOSSS spectrum \citep{sota11}, but with a $\Delta V \sim 34$ km\,s$^{-1}$. Thus, it may be an SB1.

\emph{LAMOST-O 140 = TYC 2895-2762-1}. In \citet{liu19},  its spectral type is B. But its LAMOST spectrum should be O9.7 II, and its He II $\lambda$4686 shows emission with an inverse in it. As a result, it is classified as O9.7 IIp. 

\emph{LAMOST-O 142 = HD 278 247}.  In \citet{roman19}, its spectral type is also O4 V((f)). 

\emph{LAMOST-O 143 = ALS 19710}. In \citet{liu19},  its spectral type is O8.5. But it should be O9.2 V.

\emph{LAMOST-O 145 = V* AE Aur}. This spectrum is 13''.15 away from AE Aur. By comparing with the GOSSS spectrum in \citet{sota11}, the LAMOST spectrum is well consistent with the GOSSS spectrum, but with a $\Delta V \sim 38$ km\,s$^{-1}$, which suggests it is an SB1.  

\begin{figure*}
\center
\includegraphics[angle=0, scale=0.8]{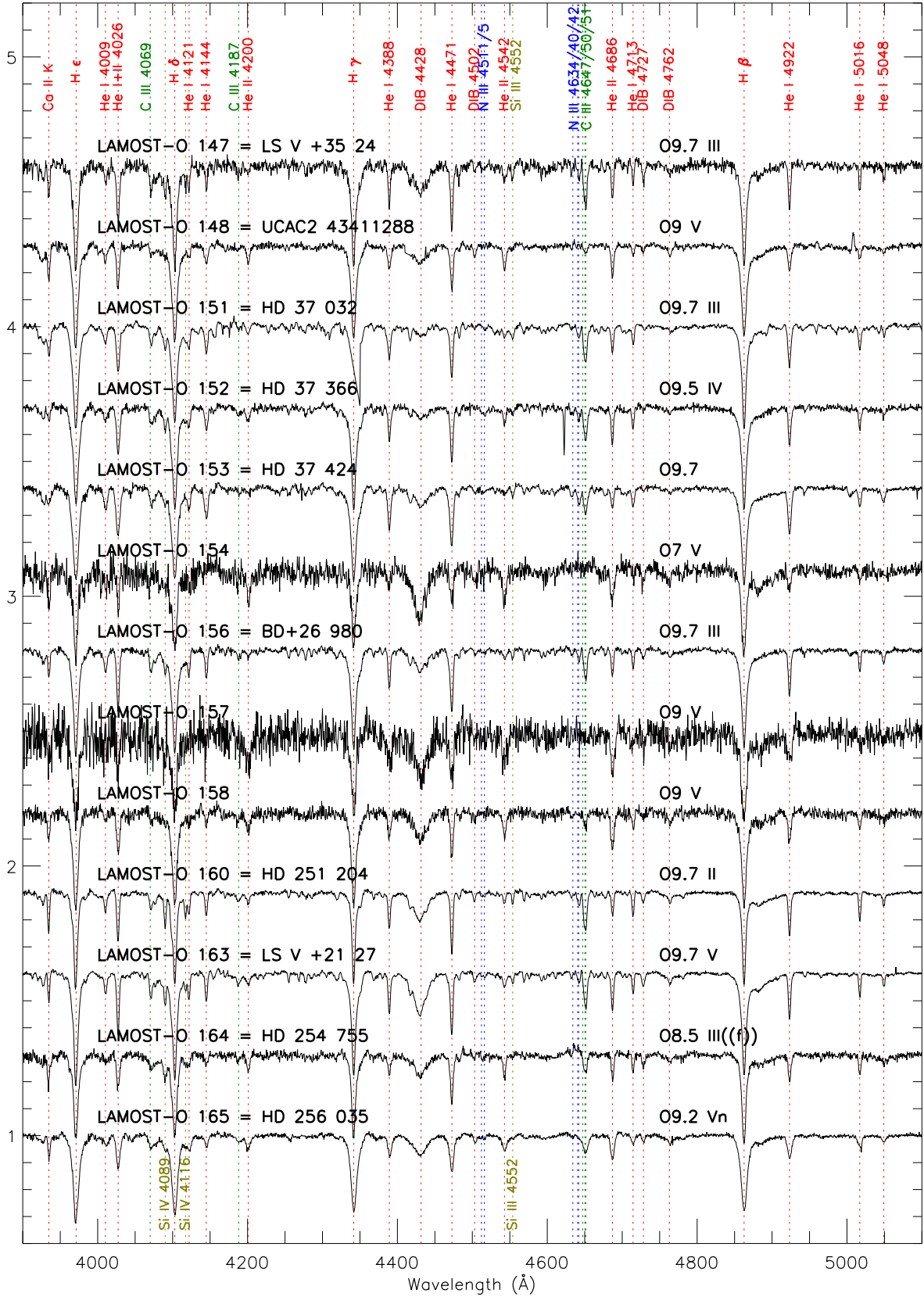}
\caption{Spectra of LAMOST-O 147, 148, 151, 152, 153, 154, 156, 157, 158, 160, 163, 164, and 165.}
\label{fig:od8}
\end{figure*}

%-------8------------
The following 13 stars are shown in Figure \ref{fig:od8}.

\emph{LAMOST-O 147 = LS V +35 24}. In \citet{geo73} and \citet{liu19}, its spectral types are O9 V. But LAMOST spectral type is O9.7 III.

\emph{LAMOST-O 148 = UCAC2 43411288}.  In \citet{roman19}, its spectral type is also O9 V.

\emph{LAMOST-O 151 = HD 37\,032}. This spectrum is 4''.27 away from HD 37\,032. In VizieR, the spectral type of  HD 37\,032 is B or OB. Its LAMOST spectrum is O9.7 III.

\emph{LAMOST-O 152 = HD 37\,366}.  This spectrum is 7''.48 away from HD 37\,366, but the LAMOST spectrum is consistent well with the GOSSS spectrum \citep{sota11}.

\emph{LAMOST-O 153 = HD 37\,424}. Its spectral type is B9 in most catalogs in VizieR, while it is B2 II in \citet{char88}. Its LAMOST spectrum is O9.7.  
The ratio of He II $\lambda$4686/He I $\lambda$4713 indicates that its luminosity is II, but its weak Si IV $\lambda$4089 and 4116 indicate it is a dwarf. Thus, its luminosity is unclear. 

\emph{LAMOST-O 154}. In \citet{liu19}, its spectral type is O, but it should be O7 V.

\emph{LAMOST-O 156 = BD+26\,980}. In \citet{roman19}, its spectral type is O9.5 V(n), but it is more like O9.7 III.

\emph{LAMOST-O 157}. In \citet{liu19}, its spectral type is O9 III. This star was mistaken for TYC 1867-418-1 by \citet{roman19},  and was assigned to O7.5 III((f)). But LAMOST spectrum shows it is an O9 V, without N III $\lambda$4636/40/42 emission.

\emph{LAMOST-O 158}. In Simbad, there is no spectral information about this star. Its LAMOST spectrum is O9 V.

\emph{LAMOST-O 160 = HD 251\,204}. In Simbad, its spectral type is B or OB. Its LAMOST spectrum is O9.7 II.

\emph{LAMOST-O 163 = LS V +21 27}. In \citet{roman19}, its spectral type is O9.5 IV, but it is more like O9.7 V.

\emph{LAMOST-O 164 = HD 254\,755}. This spectrum is 5.03'' away from HD 254\,755. In \citet{skiff14}, the spectral type of HD 254\,755 is O9 or OB. In Simbad, HD 254\,755 is a double or multiple star. Its LAMOST spectrum is O8.5 III((f)).

\emph{LAMOST-O 165 = HD 256\,035}. This spectrum is 4.62'' away from HD 256\,035. In \citet{skiff14}, the spectral type of HD 256\,035 is O9 or OB. Its LAMOST spectrum is O9.2 Vn.

%----------------9--------------

\begin{figure*}
\center
\includegraphics[angle=0, scale=0.8]{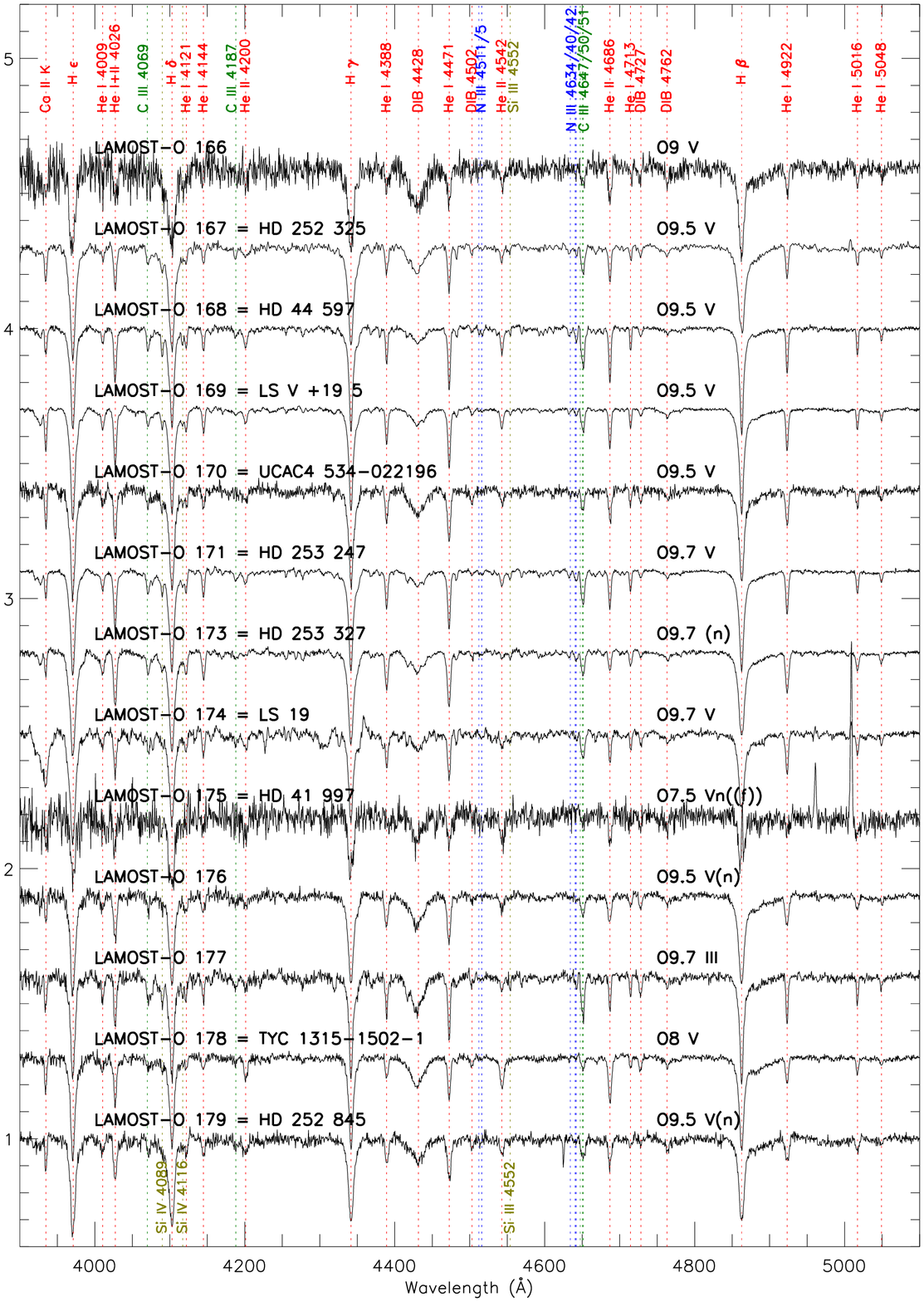}
\caption{Spectra of LAMOST-O 166, 167, 168, 169, 170, 171, 173, 174, 175, 176, 177, 178, and 179.}
\label{fig:od9}
\end{figure*}

The following 13 stars are shown in Figure \ref{fig:od9}.

\emph{LAMOST-O 166}. In \citet{liu19}, its spectral type is O9 III, but it is more like O9 V.

\emph{LAMOST-O 167 = HD 252 325}.  In \citet{roman19}, its spectral type is O9 V(n). But it is more like O9.5 V.

\emph{LAMOST-O 168 = HD 44\,597}. This spectrum is 3''.30 away from HD 44\,597.  In \citet{liu19}, its spectral type is O9 V. Here, it is O9.5 V.

\emph{LAMOST-O 169 = LS V +19 5}. In \citet{mai16}, its spectral type is also O9.5 V, but most lines are a little shallower in LAMOST spectrum.

\emph{LAMOST-O 170 = UCAC4 534-022196}. In \citet{roman19}, its spectral type is O9.5 IV((n)). Here, it is O9.5 V.

\emph{LAMOST-O 171 = HD 253\,247}. In \citet{liu19}, its spectral type is O. Here, it is O9.7 V.

\emph{LAMOST-O 173 = HD 253\,327}.  In \citet{liu19}, its spectral type is B.  Its LAMOST spectrum clearly is O9.7. However, its luminosity is unclear, because  He II $\lambda$4686/He I $\lambda$4713 suggests II, but Si IV $\lambda$4089 is very weak.

\emph{LAMOST-O 174 = LS 19}.  In \citet{ojha11}, its spectral type is O9.5 V. Its LAMOST spectrum is O9.7 V.

\emph{LAMOST-O 175 = HD 41\,997}. This spectrum is 10''.08 away from HD 41\,997. In \citet{sota11}, the spectral type of HD 41\,997 is O7.5 Vn((f)). The signal-to-noise ratio of the LAMOST spectrum is very low, but by comparing the spectra of LAMOST and GOSSS, they seem well consistent with each other. Thus, the GOSSS spectral type 7.5 Vn((f)) is adopted.

\emph{LAMOST-O 176}. In \citet{liu19}, its spectral type is O. Here, it is O9.5 V(n).

\emph{LAMOST-O 177}. In \citet{liu19}, its spectral type is B. Here, it is O9.7 III.

\emph{LAMOST-O 178 = TYC 1315-1502-1}. There is no spectral information for this star in Simbad. Its LAMOST spectrum is O8 V.

\emph{LAMOST-O 179 = HD 252\,845}.  This spectrum is 4''.14 away from HD 252\,845. In \citet{roman19}, the spectral type is O9 V:nn, but it should be O9.5 V(n).

\begin{figure*}
\center
\includegraphics[angle=0, scale=0.8]{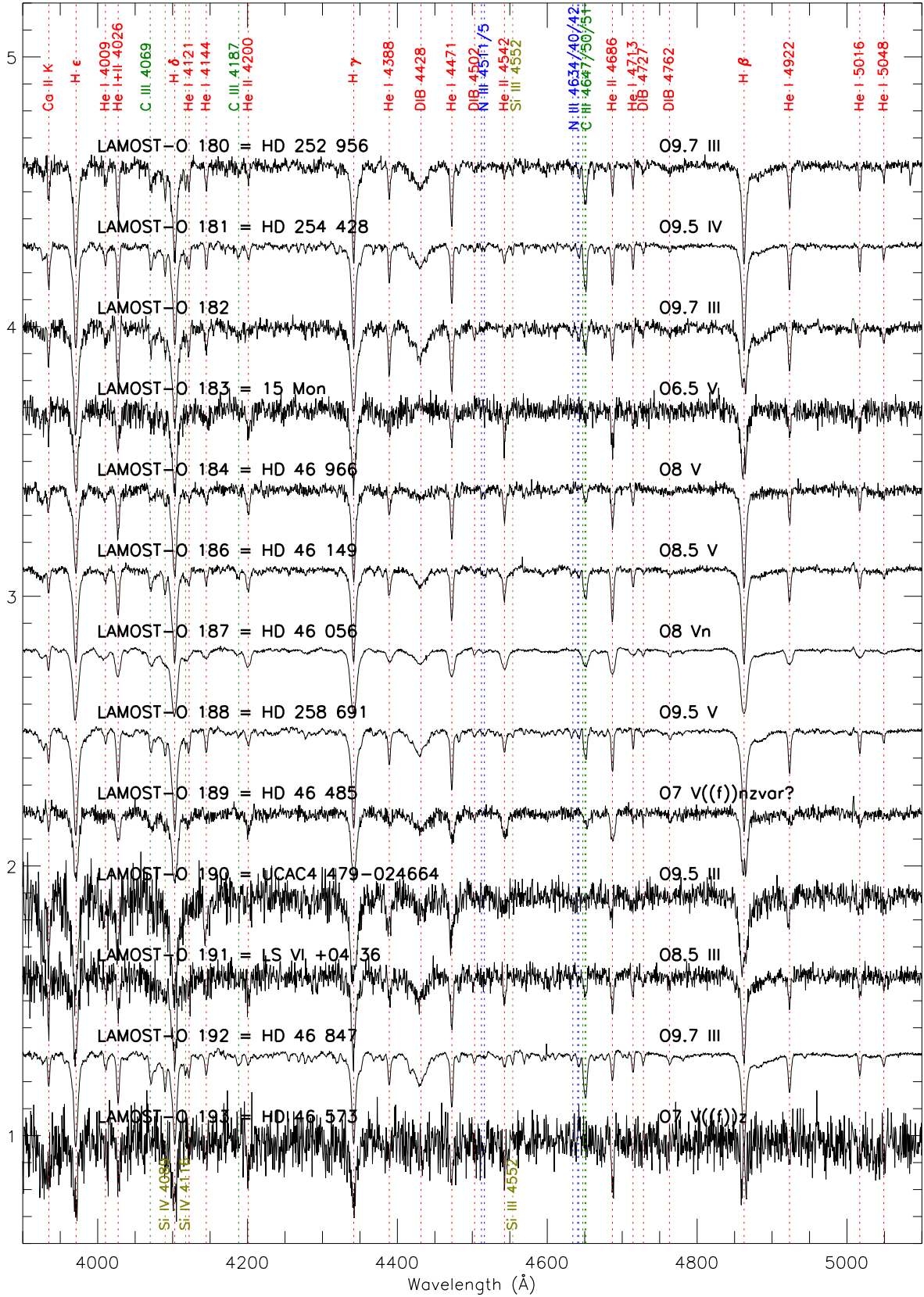}
\caption{Spectra of LAMOST-O 180, 181, 182, 183, 184, 186, 187, 188, 189, 190, 191, 192, and 193.}
\label{fig:od10}
\end{figure*}

%----------------10-------------
The following 13 stars are shown in Figure \ref{fig:od10}.

\emph{LAMOST-O 180 = HD 252\,956}.  This spectrum is 4''.42 away from HD 252\,956. In \citet{skiff14}, the spectral type is OB or B. Its LAMOST spectrum is  O9.7 III.

\emph{LAMOST-O 181 = HD 254\,428}. This spectrum is 4''.90 away from HD 254\,428. In VizieR, the spectral type is OB or around B0. Its LAMOST spectrum is O9.5 IV.

\emph{LAMOST-O 182}. In Simbad, this spectrum is 5''.50 away from ALS 18672 and 1''.66 away from the radio source NVSS J061453+122122, respectively. These three sources should be the same star in Aladin by eye inspection. In \citet{liu19}, the spectral type is B. Here, it is  O9.7 III.

\emph{LAMOST-O 183 = 15 Mon}. This spectrum is 21''.19 away from 15 Mon. In \citet{skiff14}, the spectral type is O7V + B1.5/2V. Its LAMOST spectrum seems O6.5 V.

\emph{LAMOST-O 184 = HD 46\,966}. This spectrum is 11''.54 away from HD 46\,966. This star is the O8.5 IV standard star in GOSSS. Compared with the GOSSS spectrum, He II $\lambda$4200 He II$\lambda$4388, He I $\lambda$4471, He II $\lambda$4542, and H$\gamma$ in the LAMOST spectrum become stronger. As a result, the LAMOST spectrum is more like the GOSSS O8 V standard star HD 101\,223. Thus, it is assigned O8 V.

\emph{LAMOST-O 186 = HD 46\,149}. This spectrum is 8''.95 away from HD 46\,149. This star is also in GOSSS \citep{sota11}. The spectra of LAMOST and GOSSS are well consistent with each other.

\emph{LAMOST-O 187 = HD 46\,056}. This spectrum is 6''.91 away from HD 46\,056. This star is also in GOSSS \citep{sota11}. The spectra of LAMOST and GOSSS are well consistent with each other.

\emph{LAMOST-O 188 = HD 258\,691}. In Simbad, its spectral type is O9 V.
Its LAMOST spectral type is O9.5 V.

\emph{LAMOST-O 189 = HD 46\,485}. This spectrum is 8''.77 away from HD 46\,485. This star is also in GOSSS \citep{sota14}. The signal-to-noise ratio of the LAMOST spectrum is low, but the spectra of LAMOST and GOSSS seem wekk consistent with each other. Thus, the GOSSS spectral type O7 V((f))nzvar? is adopted.

\emph{LAMOST-O 190 = UCAC4 479-024664}. In \citet{roman19}, its spectral type is O9.7 IV(n). Though low signal-to-noise ratio, it seems to be an O9.5 III.

\emph{LAMOST-O 191 = LS VI +04 36}. Its spectral type in \citet{roman19} is also O8.5 III. 

\emph{LAMOST-O 192 = HD 46\,847}. In \citet{neg04}, its spectral type is O9.7 IV. Its LAMOST spectrum is O9.7 III.

\emph{LAMOST-O 193 = HD 46\,573}. This spectrum is 11''.46 away from HD 46\,573. This star is also in GOSSS \citep{sota11}. The signal-to-noise ratio of the LAMOST spectrum is low, but the spectra of LAMOST and GOSSS seem consistent with each other. Thus, the GOSSS spectral type O7 V((f))z is adopted.

\begin{figure*}
\center
\includegraphics[angle=0, scale=0.8]{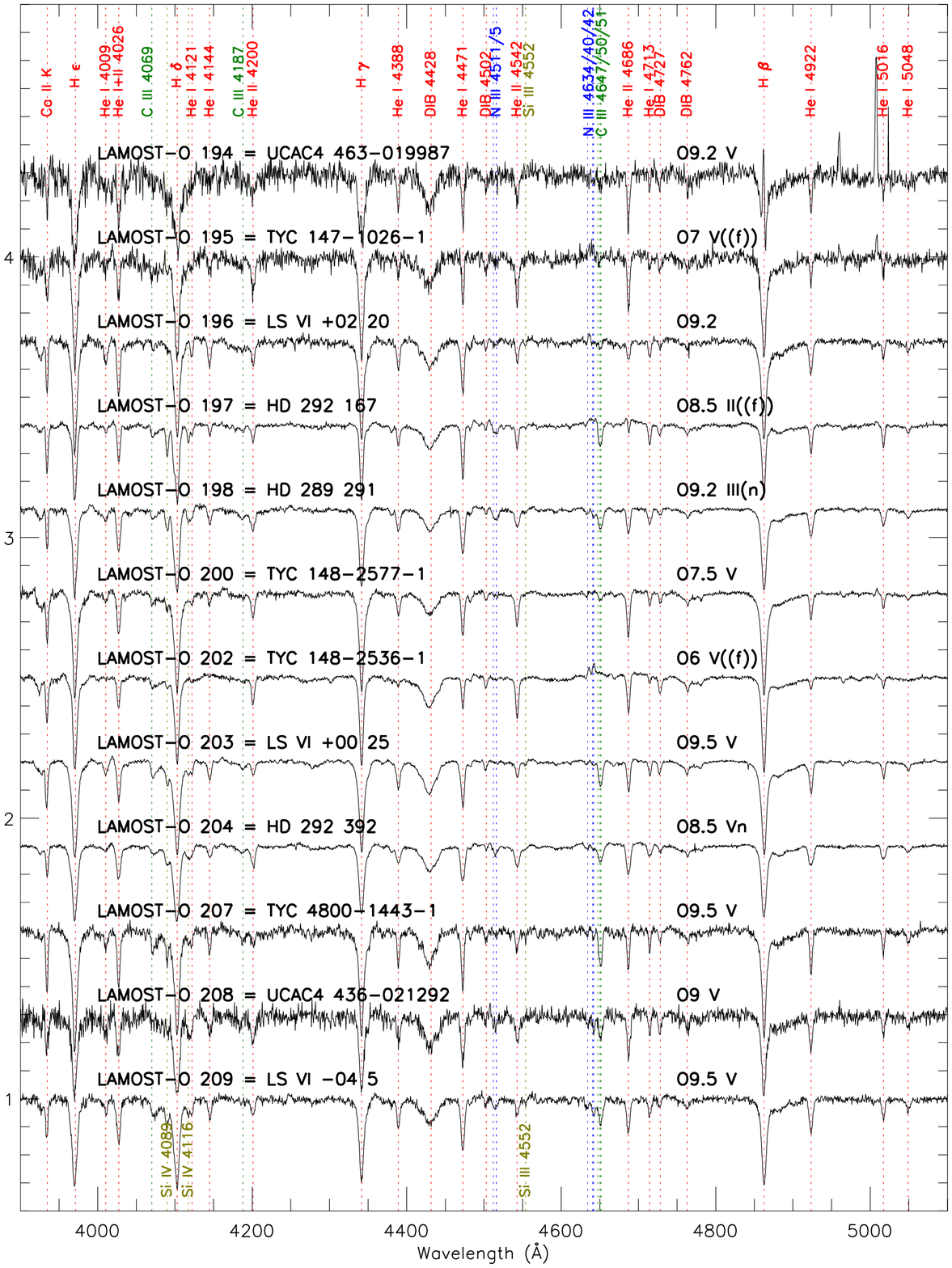}
\caption{Spectra of LAMOST-O 194, 195, 196, 197, 198, 200, 202, 203, 204, 207, 208, and 209.}
\label{fig:od11}
\end{figure*}

%----------------11-------------
The following 10 stars are shown in Figure \ref{fig:od11}.

\emph{LAMOST-O 194 =UCAC4 463-019987}.  In \citet{roman19}, its spectral type is O8 V((f))e. Here, it is O9.2 V. 

\emph{LAMOST-O 195 = TYC 147-1026-1}. In \citet{rus07}, its spectral type is O7 V. Its LAMOST spectrum is O7 V((f)).

\emph{LAMOST-O 196 = LS VI +02 20}. In \citet{nas65}, its spectral type is OB. Its LAMOST spectrum is O9.2. However, its luminosity is unclear, because  He II $\lambda$4686/He I $\lambda$4713 suggests II, but there are not Si IV $\lambda$4089 and 4116.

\emph{LAMOST-O 197 = HD 292\,167}. In \citet{neg15}, its spectral type is O8.5 Ib(f). The emission of N III $\lambda$4634/40/42 is very weak in the LAMOST spectrum, while Si IV $\lambda$4089 and 4116 are not strong enough to be Ib, so its spectral type is O8.5 II((f)).

\emph{LAMOST-O 198 = HD 289\,291}. In \citet{hilt56}, its spectral type is O8:. Its LAMOST spectrum is O9.2 III(n).

\emph{LAMOST-O 200 = TYC 148-2577-1}. In \citet{seb12}, its spectral type is B1 V. Its LAMOST spectral type is O7.5 V.

\emph{LAMOST-O 202 = TYC 148-2536-1}.  In \citet{neg15}, its spectral type is O6 V((f)). Its LAMOST spectral type is O6.5 V((f)).

\emph{LAMOST-O 203 = LS VI +00 25} In \citet{mun99}, it is an O7 V SB1, with a period of 11.030 days. Its LAMOST spectrum is O9.5 V.

\emph{LAMOST-O 204 = HD 292\,392}.  In \citet{neg15}, its spectral type is also O8.5 V. Moreover, its LAMOST spectrum shows it is also a fast rotator, thus an O8.5 Vn star.

\emph{LAMOST-O 207 = TYC 4800-1443-1}. In \citet{mcc56}, its spectral type is B0:. Its LAMOST spectrum is O9.5 V.

\emph{LAMOST-O 208 = UCAC4 436-021292}. In \citet{seb12}, its spectral type is B0.5 Ia. Its LAMOST spectrum is O9 V.

\emph{LAMOST-O 209 = LS VI -04 5}. In \citet{nas65}, its spectral type is OB. Its LAMOST spectrum is O9.5 V.

\section{Summary} \label{sec:sum}

This paper publishes 209 O-type stars with their spectra found in the LAMOST database up to 2018. Ninety-four new stars are firstly given. The spectral classification standards are from \citet{sota11, sota14} and \citet{mai16}. All spectra in this paper are compared with standard stars in Table 2 in \citet{mai16} to determine spectral types. But there are still many stars that cannot be assigned correct spectral types because of the influence of their company and/or other unknown reasons.
\par

 For a star that has multiple LAMOST spectra from  different epochs, or has both LAMOST and GOSSS spectra, all these spectra are compared to inspect spectral variations or/and wavelength shifts to find binaries. 
\par
All LAMOST low- and medium-resolution spectra shown in this paper are provided in the China-VO PaperData repository: 
\href{https://nadc.china-vo.org/registry/paperdata/101049}{doi:10.12149/101049}

\acknowledgments

The author thanks the anonymous referee for very helpful comments. 
This research is supported by the National Natural Science Foundation of China (NSFC; grant No. 12073038).  The author acknowledges the support from Cultivation Project for LAMOST Scientific Payoff and Research Achievement of CAMS-CAS. The author acknowledges the support from the Chinese Space Station Telescope (CSST) project.
\par
Guoshoujing Telescope (the Large Sky Area Multi-Object Fiber Spectroscopic Telescope LAMOST) is a National Major Scientific Project built by the Chinese Academy of Sciences. Funding for the project has been provided by the National Development and Reform Commission. LAMOST is operated and managed by the National Astronomical Observatories, Chinese Academy of Sciences.


\vspace{5mm}
\facilities{LAMOST}


%\software{astropy \citep{2013A&A...558A..33A},  
%          Cloudy \citep{2013RMxAA..49..137F}, 
 %         SExtractor \citep{1996A&AS..117..393B}
  %        }



\appendix

\startlongtable
\centerwidetable
\begin{deluxetable}{cccccccc}
\tablenum{1}
\tablecaption{Spectral Information for 209 LAMOST O-type Stars\label{tab:lamosto}}
\tablewidth{0pt}
\tablehead{
\colhead{LAMOST Name} & \colhead{R. A.(deg)} & \colhead{ Decl.(deg)} &\colhead{Simbad Name} & \colhead{Dist (arcsec)} & \colhead{SpT} & \colhead{Mult} & \colhead{New} 
}
%\decimalcolnumbers
\startdata
LAMOST-O 001&279.765732&-7.859846&HD 172 175& &O6 I(n)fp& & \\
LAMOST-O 002&280.401734&-5.956290&TYC 5125-2083-1& &O9 II& &T\\
LAMOST-O 003&281.144639&-5.939243&TYC 5125-3040-1& &O + O&SB2&T\\
LAMOST-O 004&281.894538&-6.307544&HD 173 820& &O8.5 V& &T\\
LAMOST-O 005&282.126703&-5.519835&BD-05 4769& &O7.5 III((f))& & \\
LAMOST-O 006&281.251027&-4.508252& & &O7.5 II((f))& &T\\
LAMOST-O 007&281.357364&-4.016850&LS IV -04 20& &O9.7 II-IIIn& &T\\
LAMOST-O 008&281.233639&-3.812636&TYC 5121-769-1& &O8 Ib((f))& &T\\
LAMOST-O 009&282.456935&-4.327860&BD-04 4593& &O9.7 II& & \\
LAMOST-O 010&282.608507&-4.248241&TYC 5122-465-1& &O9 IIIne& &T\\
LAMOST-O 011&281.709165&-3.730182&ALS 19303& &O7 V& &T\\
LAMOST-O 012&277.755725&13.503585&TYC 1036-450-1& &O9.7&SB1& \\
LAMOST-O 013&295.051991&23.627912&LS II +23 14& &O9.7 IIIne& &B\\
LAMOST-O 014&300.340211&22.291709&HD 345 475& &O9.7 II& &T\\
LAMOST-O 015&294.743911&26.145659& & &O9.7 II& &T\\
LAMOST-O 016&296.425477&25.354556&HD 338 916& &O7.5 Vz& & \\
LAMOST-O 017&300.825136&30.651276& & &O9.5 V& &T\\
LAMOST-O 018&305.805612&37.286904&[NH52] 74& &O9.7 IV& &B\\
LAMOST-O 019&305.927422&37.899148&BD+37 3917& &O9.7 II& &T\\
LAMOST-O 020&304.948209&38.595413&HD 228 943& &O9.7 III(n)&SB1& \\
LAMOST-O 021&306.029716&38.022734&TYC 3152-1390-1& &O9.7 IV& &T\\
LAMOST-O 022&305.586713&38.724021&HD 194 094& &O9 III& & \\
LAMOST-O 023&307.398459&40.408733&TYC 3156-1881-1& &O7 Vz&SB1& \\
LAMOST-O 024&307.406578&40.852362&UCAC4 655-091969& &O8 III((f))& &B\\
LAMOST-O 025&308.160176&40.679025&[CPR2002] A25& &O8 III& & \\
LAMOST-O 026&307.740527&41.165973&[CPR2002] A26& &O9 III& & \\
LAMOST-O 027&308.295679&41.252285&Schulte 9& &O4.5 If&SB1& \\
LAMOST-O 028&307.665858&41.614064&GSC 03161-01264& &O6.5 Vz& & \\
LAMOST-O 029&307.956907&41.474032&Schulte 20& &O9.7 IV&SB1& \\
LAMOST-O 030&308.404145&41.269810&Schulte 70& &O9.5 III(n)&SB1& \\
LAMOST-O 031&308.208404&41.395749&Schulte 17& &O8 V&SB1& \\
LAMOST-O 032&308.420264&41.505266&GSC 03161-01086& &O7 Vz& & \\
LAMOST-O 033&308.443380&41.550192&BD+41 3804& &O9.7 Iab& & \\
LAMOST-O 034&308.556298&41.584122&Schulte 29& &O7 V(n)((f))z& & \\
LAMOST-O 035&308.741030&41.731738&RLP 1252& &O8 Ve& &B\\
LAMOST-O 036&306.576874&43.796454&UCAC4 669-086327& &O4 IV(fc)p& &B\\
LAMOST-O 037&308.842778&43.925132&2MASS J20352227+4355304& &O9.7 II& &B\\
LAMOST-O 038&331.984167&54.518333&BD+53 2790& &O9.2 ne& & \\
LAMOST-O 039&334.491164&54.467304&V* AW Lac& &B n + O9.5 Vn&SB2&T\\
LAMOST-O 040&335.560943&54.800355&HD 235 813& &O9.7 II& &T\\
LAMOST-O 041&337.475207&55.407962&BD+54 2789& &O9.2 V + B&SB2&T\\
LAMOST-O 042&343.384265&52.638135&HD 235 989& &O9.2 IIInn& &T\\
LAMOST-O 043&343.938150&56.476847&LS III +56 109& &O5.5 V(n)((fc))&SB1& \\
LAMOST-O 044&343.765772&57.620883&LS III +57 99& &O8.5 Vn + O8.5 Vn&SB2&T\\
LAMOST-O 045&344.807124&58.745279&LS III +58 70& &O9.7 IV& &T\\
LAMOST-O 046&343.422151&60.318421& & &O9 II& &T\\
LAMOST-O 047&347.142318&56.016970&HD 240 197& &ON9.7 (n)e& &T\\
LAMOST-O 048&345.028540&59.324992&BD+58 2520& &O9.2 V& &T\\
LAMOST-O 049&346.952327&61.162029& & &O8 V((f))& &T\\
LAMOST-O 050&348.455916&59.640248& & &O9 V& &T\\
LAMOST-O 051&348.648781&59.553593&Hilt 1202& &O9.7 IIn& &T\\
LAMOST-O 052&348.128234&60.215052& & &O9.5 V& &T\\
LAMOST-O 053&348.643591&59.836733&HD 240 234& &O9.7 V(n)e& &T\\
LAMOST-O 054&348.701002&59.921967&LS III +59 58& &O9.5 Vn& &T\\
LAMOST-O 055&349.080820&59.455806&LS III +59 63& &O7 III(n)((f))& &T\\
LAMOST-O 056&349.148451&59.764180& & &O9.5 V + O9.5 V&SB2&T\\
LAMOST-O 057&349.032066&60.042411&TYC 4279-1192-1& &O9.5 V& & \\
LAMOST-O 058&349.273683&59.912731&IRAS 23149+5938& &O9.7 III& &T\\
LAMOST-O 059&349.304273&60.107976&TYC 4279-982-1& &O9 Vn& &T\\
LAMOST-O 060&348.954883&61.133210& & &O8 V + B&SB2& \\
LAMOST-O 061&353.198135&59.951567&LS I +59 5& &O7.5 V((f))& &T\\
LAMOST-O 062&354.175239&58.657754&LS III +58 86& &O7 V(n)z& &T\\
LAMOST-O 063&353.236693&60.662131&LS I +60 8& &O9.5 III& & \\
LAMOST-O 064&353.405331&60.558938&LS I +60 9& &O7 Vn((f))z& &T\\
LAMOST-O 065&353.810688&60.006449&EM* GGR 149& &O9.7 V(n)e& &B\\
LAMOST-O 066&353.403528&60.752068& & &O9.5 V& & \\
LAMOST-O 067&353.385451&60.792254& & &O9: V& & \\
LAMOST-O 068&353.408224&61.079876&LS I +60 10& &O7.5 Vz& &T\\
LAMOST-O 069&354.964232&60.091374&LS I +59 11& &O7.5 III((f))& &T\\
LAMOST-O 070&355.758156&59.065586&BD+58 2636& &O9.2 III& &T\\
LAMOST-O 071&355.546938&60.943279&TYC 4280-103-1& &O9.7 IV& &T\\
LAMOST-O 072&356.403122&59.906001&V* QQ Cas& &O8 Vn + O9 Vn&SB2&T\\
LAMOST-O 073&354.909499&62.666785&LS I +62 4& &O9.5 V&SB1&T\\
LAMOST-O 074&355.944044&61.450150&LS I +61 28& &ON8.5 Vn& &B\\
LAMOST-O 075&356.519854&61.783478&LS I +61 50& &O7.5 Vn& &T\\
LAMOST-O 076&356.369232&63.271331&LS I +62 14& &O5: Iafpe& &T\\
LAMOST-O 077&358.248421&60.481672&LS I +60 50& &O9.5 III& & \\
LAMOST-O 078&358.169361&60.993612&BD+60 2632& &O9.5 IV& &T\\
LAMOST-O 079&358.351876&60.680337&HD 240 435& &O9.7 III& &T\\
LAMOST-O 080&358.271744&60.912417&BD+60 2635& &O6 V((f))&SB1& \\
LAMOST-O 081&357.221270&63.005322&LS I +62 25& &O9.7 III&SB1&T\\
LAMOST-O 082&357.018620&63.397473&BD+62 2299& &O8 II((f))& &T\\
LAMOST-O 083&359.296667&60.485096&LS I +60 62& &O8.5 V& &T\\
LAMOST-O 084&359.712130&60.264174&HD 240 464& &O9.5 V&SB1& \\
LAMOST-O 085&18.648234&57.924769&HD 236 672& &ON9 IVn& &B\\
LAMOST-O 086&32.375302&58.783764&HD 13 022& &O9.7 II-III&SB1& \\
LAMOST-O 087&33.124871&59.901163&Hilt 233& &O8 V + B0: V&SB2& \\
LAMOST-O 088&35.687196&61.839935&LS I +61 266& &O6 Vz& &T\\
LAMOST-O 089&36.150672&61.687750&LS I +61 267& &O7 V(n)& &T\\
LAMOST-O 090&31.627886&54.519031& & &O9.7 V& &T\\
LAMOST-O 091&37.578910&63.289281&LS I +63 187& &O9.7 III& &T\\
LAMOST-O 092&36.643315&62.011796&BD+61 411& &O7 V((f))z&SB1& \\
LAMOST-O 093&38.798195&63.584176&BD+62 419& &O9.7 III& &T\\
LAMOST-O 094&35.544578&59.549696&HD 14 442& &O5 Infp&SB1& \\
LAMOST-O 095&37.376963&61.495595&V* KM Cas& &O9 Vne& &B\\
LAMOST-O 096&39.075907&62.948151&BD+62 424& &O6 V((f))&SB1& \\
LAMOST-O 097&38.045221&61.552210&BD+60 498& &O9.7 IV&SB1& \\
LAMOST-O 098&38.069798&61.554168&BD+60 499& &O9.5 V&SB1& \\
LAMOST-O 099&38.151151&61.473773&BD+60 501& &O7 V(n)((f))z&SB1& \\
LAMOST-O 100&36.004714&58.323855&HD 14 645& &O9.7 IIIne& &T\\
LAMOST-O 101&35.799014&57.472052&BD+56 594& &O7.5 Vz& &T\\
LAMOST-O 102&38.309007&60.022766&LS I +59 139& &O9.7 IV& &T\\
LAMOST-O 103&38.059327&59.621146&TYC 3699-1537-1& &O9.7 IV& &T\\
LAMOST-O 104&41.294227&62.890463&LS I +62 223& &O6.5 Vz + O9 V&SB2&T\\
LAMOST-O 105&40.284160&61.135197&BD+60 544& &O9.7 III&SB1&T\\
LAMOST-O 106&42.250807&62.535328&Lan 27& &O9.2 V& &T\\
LAMOST-O 107&40.836558&60.836840& & &O9.7 V& &T\\
LAMOST-O 108&42.213030&62.216246&LS I +62 227& &O9.7 Ib + O9 V&SB2&T\\
LAMOST-O 110&42.884097&62.222594& & &O9.5 IV& & \\
LAMOST-O 111&41.542071&60.266033&2MASS J02461010+6015576& &O& &T\\
LAMOST-O 112&42.276544&60.560702&LS I +60 263& &O7 III((f))& &T\\
LAMOST-O 113&44.267202&61.416018&BD+60 594& &O8.5 Vn&SB1& \\
LAMOST-O 114&35.715712&41.480129&HD 14 633&28.34&O& & \\
LAMOST-O 115&50.872389&55.206651&LS I +55 47& &O9.5 IV& & \\
LAMOST-O 116&59.170522&57.257971&LS I +57 136& &O9.2 V& & \\
LAMOST-O 117&59.870331&57.118007&BD+56 866& &O9.2 V(n)& & \\
LAMOST-O 118&57.217761&53.149000& & &O& &T\\
LAMOST-O 119&61.244651&55.909252&LS V +55 12& &O9.7 II& &T\\
LAMOST-O 120&61.841915&55.507740&TYC 3722-435-1& &O9.2 V& &T\\
LAMOST-O 121&61.682058&54.396626&LAMOST J040643.69+542347.8& &O6.5 Vnnn((f))p& &B\\
LAMOST-O 122&60.836444&51.314594&BD+50 886& &O4 V((c))&SB1& \\
LAMOST-O 123&65.255647&53.609539&LS V +53 20& &O9.2 V& &B\\
LAMOST-O 124&61.470965&51.116128&TYC 3339-851-1& &O6 Vz& &B\\
LAMOST-O 125&63.074008&52.030896&LS V +51 16& &O9.7 V& &B\\
LAMOST-O 126&64.964828&53.158056&GSC 03719-00546& &O9.7 V& & \\
LAMOST-O 127&64.649308&52.865094&BD+52 805& &O9 V&SB1& \\
LAMOST-O 128&65.292516&53.171164&LS V +53 21& &O9.7 II& &T\\
LAMOST-O 129&65.353371&52.950100&TYC 3719-1248-1& &O9.7 IV(n)& &T\\
LAMOST-O 130&62.794727&50.708221&TYC 3340-2437-1& &O9.7 III&SB1&T\\
LAMOST-O 131&59.715117&48.301529&LS V +48 9& &O9.7 n&SB1&B\\
LAMOST-O 132&65.191785&51.893107&LS V +51 18& &O7.5 V& &T\\
LAMOST-O 133&67.748920&52.664967&TYC 3732-701-1& &O9.7 III& &T\\
LAMOST-O 134&67.653280&52.581847&TYC 3732-745-1& &O9.7 III(n)& &T\\
LAMOST-O 135&63.833789&48.513409&LS V +48 12& &O9.5 Binary&SB2&B\\
LAMOST-O 136&59.740276&35.798951&ksi Per&28.65&O7.5 III(n)((f))&SB1& \\
LAMOST-O 137&72.580300&45.013304&UCAC4 676-031103& &O9.5 III& &B\\
LAMOST-O 138&91.467910&48.247255&HD 41 161&7.48&O8 Vn&SB1& \\
LAMOST-O 139&79.594814&41.935024&HD 277 878& &O7 V((f))z&SB1&B\\
LAMOST-O 140&77.457310&37.586155&TYC 2895-2762-1& &O9.7 IIp& &T\\
LAMOST-O 141&80.002659&38.912102&LS V +38 12& &O6.5 V((f)) + B0 III-V&SB2& \\
LAMOST-O 142&81.810394&39.852325&HD 278 247& &O4 V((f))& &B\\
LAMOST-O 143&78.358283&37.446234&ALS 19710& &O9.2 V& &B\\
LAMOST-O 144&82.521290&38.440365&LS V +38 15& &O9 V + B&SB2&B\\
LAMOST-O 145&79.072462&34.314875&V* AE Aur&13.15&O9.5 V&SB1& \\
LAMOST-O 146&81.889888&34.450499&LS V +34 21& &O9 IV&SB1&B\\
LAMOST-O 147&84.940781&35.898988&LS V +35 24& &O9.7 III& & \\
LAMOST-O 148&80.692431&33.422328&UCAC2 43411288& &O9 V& &B\\
LAMOST-O 149&80.665368&33.371715&LS V +33 15& &O5.5 Vz((f)) var&SB1& \\
LAMOST-O 150&80.666955&33.319217&HD 242 926& &O7 Vz& & \\
LAMOST-O 151&84.288511&34.811139&HD 37 032&4.27&O9.7 III& &T\\
LAMOST-O 152&84.853588&30.892837&HD 37 366&7.48&O9.5 IV& & \\
LAMOST-O 153&84.934992&27.780888&HD 37 424& &O9.7& &T\\
LAMOST-O 154&87.241714&27.352786& & &O7 V& &B\\
LAMOST-O 155&91.741272&29.058012&LS V +29 22& &O9.7&SB1&T\\
LAMOST-O 156&87.980878&27.032787&BD+26 980& &O9.7 III& &B\\
LAMOST-O 157&88.059307&25.125726& & &O9 V& &B\\
LAMOST-O 158&90.290288&25.595077& & &O9 V& &T\\
LAMOST-O 159&96.249449&26.822050&RL 128& &O4.5 V((c))z& &B\\
LAMOST-O 160&91.273607&23.394041&HD 251 204& &O9.7 II& &T\\
LAMOST-O 161&94.923153&23.289703&HD 255 055& &O9.2 IVe& &B\\
LAMOST-O 162&95.135680&23.317320&HD 255 312& &O6 Vzn + O9 Vn&SB2&T\\
LAMOST-O 163&92.133750&21.610911&LS V +21 27& &O9.7 V& &B\\
LAMOST-O 164&94.633763&22.678598&HD 254 755&5.03&O8.5 III((f))& & \\
LAMOST-O 165&95.744063&22.862915&HD 256 035&4.62&O9.2 Vn& & \\
LAMOST-O 166&91.091069&20.273966& & &O9 V& &B\\
LAMOST-O 167&92.251298&20.640514&HD 252 325& &O9.5 V& &B\\
LAMOST-O 168&95.869889&20.392120&HD 44 597&3.30&O9.5 V& &B\\
LAMOST-O 169&96.257907&19.848470&LS V +19 5& &O9.5 V& & \\
LAMOST-O 170&90.245143&16.668620&UCAC4 534-022196& &O9.5 V& &B\\
LAMOST-O 171&93.092140&18.016080&HD 253 247& &O9.7 V& &B\\
LAMOST-O 172&95.427815&19.160047&TYC 1323-1592-1& &O8.5 Vz& &B\\
LAMOST-O 173&93.184080&17.987310&HD 253 327& &O9.7 (n)& &T\\
LAMOST-O 174&93.267560&17.978180&LS 19& &O9.7 V& & \\
LAMOST-O 175&92.230628&15.702989&HD 41 997&10.08&O7.5 Vn((f))& & \\
LAMOST-O 176&94.637895&15.279759& & &O9.5 V(n)& &B\\
LAMOST-O 177&94.564093&15.131532& & &O9.7 III& &T\\
LAMOST-O 178&94.889682&15.293595&TYC 1315-1502-1& &O8 V& &T\\
LAMOST-O 179&92.700002&13.183337&HD 252 845&4.14&O9.5 V(n)& &B\\
LAMOST-O 180&92.776887&13.145328&HD 252 956&4.42&O9.7 III& &T\\
LAMOST-O 181&94.225125&13.508074&HD 254 428&4.49&O9.5 IV& &T\\
LAMOST-O 182&93.722302&12.356639& &5.50&O9.7 III& &T\\
LAMOST-O 183&100.239310&9.892698&15 Mon&21.19&O6.5 V& & \\
LAMOST-O 184&99.107929&6.086396&HD 46 966&11.54&O8 V& & \\
LAMOST-O 185&100.500080&6.342522&HD 48 099&15.73&O5 V((f))z + O9:V&SB2& \\
LAMOST-O 186&97.970212&5.035209&HD 46 149&8.95&O8.5 V& & \\
LAMOST-O 187&97.835875&4.832788&HD 46 056&6.91&O8 Vn& & \\
LAMOST-O 188&97.638789&4.690992&HD 258 691& &O9.5 V& & \\
LAMOST-O 189&98.461242&4.527636&HD 46 485&8.77&O7 V((f))nzvar?& & \\
LAMOST-O 190&101.434850&5.736885&UCAC4 479-024664& &O9.5 III& &B\\
LAMOST-O 191&102.041760&4.821117&LS VI +04 36& &O8.5 III& &B\\
LAMOST-O 192&98.932530&2.711233&HD 46 847& &O9.7 III& & \\
LAMOST-O 193&98.597772&2.537304&HD 46 573&11.46&O7 V((f))z& & \\
LAMOST-O 194&98.646611&2.456512&UCAC4 463-019987& &O9.2 V& &B\\
LAMOST-O 195&99.557164&0.736640&TYC 147-1026-1& &O7 V((f))& & \\
LAMOST-O 196&102.763422&2.098965&LS VI +02 20& &O9.2& &T\\
LAMOST-O 197&101.225133&0.620320&HD 292 167& &O8.5 II((f))& & \\
LAMOST-O 198&102.993561&1.376179&HD 289 291& &O9.2 III(n)& & \\
LAMOST-O 199&101.167267&0.255560&TYC 147-1646-1& &O9.7&SB1&T\\
LAMOST-O 200&101.259080&0.224860&TYC 148-2577-1& &O7.5 V& &T\\
LAMOST-O 201&101.269638&0.229797&Cl* Dolidze 25 MV 17& &O7 Vz& & \\
LAMOST-O 202&101.350887&0.229642&TYC 148-2536-1& &O6 V((f))& & \\
LAMOST-O 203&102.210323&0.377128&LS VI +00 25& &O9.5 V& & \\
LAMOST-O 204&102.573440&0.446564&HD 292 392& &O8.5 Vn& & \\
LAMOST-O 205&102.181212&-0.257652&HD 292 419& &O6 Iafp& &T\\
LAMOST-O 206&103.183444&-0.187991&EM* RJHA 83& &O9.5 Vne& &B\\
LAMOST-O 207&102.596035&-0.540200&TYC 4800-1443-1& &O9.5 V& &T\\
LAMOST-O 208&101.550781&-2.914206&UCAC4 436-021292& &O9 V& &T\\
LAMOST-O 209&99.753638&-4.380793&LS VI -04 5& &O9.5 V& &T\\
\enddata
\tablecomments{(1) In the last column, 'T' denotes the new O-type stars found with LAMOST firstly given in this work, while 'B' denotes the new ones given in \citet{liu19}, \citet{roman19}, \citet{li20a}, and \citet{li20b}; (2) the positions given in this table are LAMOST fiber positions, which may be not the exact positions of stars with $V < 10$ mag, because bright stars may saturate spectra \citep{lei14}. As a compromise, LAMOST fibers were positioned to their neighbor's places. The distances between fiber positions and star positions are given in the fifth column in parseconds, if any; (3) LAMOST-O 109 is an SB2, which will be published in a future paper.}
\end{deluxetable}
%\end{longtable}



\end{document}